\documentclass[12pt]{article}
\usepackage{graphics}
\usepackage{epsfig}
\usepackage{amsfonts}
\usepackage{booktabs}
\usepackage{graphicx}
\usepackage{color}
\usepackage{url}
\usepackage{theorem}
\usepackage{amsmath}
\usepackage{multirow}
\usepackage{bm}
\usepackage{lineno}
\usepackage{setspace}

\usepackage[utf8]{inputenc}
\usepackage[english]{babel}

\usepackage{natbib}
\usepackage{hyperref}
\usepackage{color}

\def\max{\mathop{\rm max}\limits}

\def\varY{{\cal Y}}
\def\yit{y_{it}}

\def\yonedot{\mathbf{y}_{1\boldsymbol{\cdot}}}
\def\yidot{\mathbf{y}_{i\boldsymbol{\cdot}}}
\def\yNdot{\mathbf{y}_{N\boldsymbol{\cdot}}}

\def\varX{{\cal X}}
\def\xit{\mathbf{x}_{it}}
\def\txit{\tilde{\mathbf{x}}_{it}}

\def\Xonedot{\mathbf{X}_{1\boldsymbol{\cdot}}}
\def\Xjdot{\mathbf{X}_{j\boldsymbol{\cdot}}}
\def\Xpdot{\mathbf{X}_{p\boldsymbol{\cdot}}}

\def\Xmj{\mathbb{X}_{-j}}

\def\bfbetat{\boldsymbol{\beta}_{t}}
\def\bfbeta{\boldsymbol{\beta}}

\def\bftbeta{\tilde{\boldsymbol{\beta}}}
\def\bftbetat{\tilde{\boldsymbol{\beta}}_t}

\def\Mk{{\cal M}_k}

\newtheorem{theorem}{Theorem}

\begin{document}

\baselineskip  32pt
\doublespacing
\title{Ultra High Dimensional Change Point Detection }

\author{
  Xin~Liu,~ Liwen Zhang,~Zhen~Zhang\\
  School of Statistics and Management\\
  Shanghai University of Finance and Economics\\
  Shanghai, China, 200433 \\
}

\maketitle

\begin{abstract}
Structural breaks have been commonly seen in applications. Specifically for detection of change points in time, research gap still remains on the setting in ultra high dimension, where the covariates may bear spurious correlations. In this paper, we propose a two-stage approach to detect change points in ultra high dimension, by firstly proposing the dynamic titled current correlation screening method to reduce the input dimension, and then detecting possible change points in the framework of group variable selection. Not only the spurious correlation between ultra-high dimensional covariates is taken into consideration in variable screening, but non-convex penalties are studied in change point detection in the ultra high dimension. Asymptotic properties are derived to guarantee the asymptotic consistency of the selection procedure, and the numerical investigations show the promising performance of the proposed approach.

\end{abstract}

\section{Introduction}\label{sec:intro}
Detection of structural changes in panel data models has been of particular interest since last two decades and a vast amount of literature has been developed; see, for example, \cite{csorgo1997limit,bai1998estimating,lee2011testing,li2015hysteretic,Barigozzi2018,FisherJensen2019}. A typical way detecting the unknown number of structural breaks are frequently transformed into a variable selection problem, where only a small portion of them are believed as relevant. 

Thanks to the development of high-dimensional regression methods, change point detection appears feasible with moderate number of covariates under this framework. For instance, \citep{Li2016Panel} introduced a penalized principal component (PPC) estimation procedure with an adaptive group fused LASSO to detect the multiple structural breaks in panel data models with unobservable interactive fixed effects. \citep{Qian2016Shrinkage} proposed penalized generalized method of moments (PGMM) and penalized least squares (PLS) method to determine the number of structural changes with endogenous regressors. \citep{Qian2016multiple} considered estimation and inference of common breaks in multiple panel linear regression models via adaptive group fused LASSO (AGF-LASSO). 
\citep{MaSu2018} studied the estimation of a large dimensional factor model when the factor loadings exhibit an unknown number of changes over time by the adaptive fused group Lasso procedure.
\citep{SuWang2019} proposed a heterogeneous time-varying panel data model with a latent group structure that allows varying coefficients over both individuals and time. However, data with thousands even millions of variables are rarely astounding in recent research areas including oncology images, genomics and financial time series, where usually the dimension of the observed covariates is much greater than the sample size. Consequently, two concerns come up when identifying change points under the standard framework.

A main problem turns out that 
the covariates are usually correlated in the ultra-high dimension. Accordingly, variable screening seems imperative to extract relevant covariates carrying truly useful information in data, before other statistical models are formally built and statistical inference is conducted. Much effort is paid to the investigation of variable screening in the last decades. A main technique is to consider the importance from each covariate marginally based on some specific metric. For example, \citep{Fan2008} proposed the \textit{sure independent screening} (SIS) method in ultra-high dimensional linear models, which ranks the importance of the covariates by individual marginal correlation with the response and determines the relevant covariates with larger correlations into the model controlled by a threshold. Similar idea was further extended to generalized linear models (GLM) and additive models with ultra high dimensions either by maximum marginal likelihood estimates or marginal nonparametric estimates \citep{fan2010, fan2011}. A forward regression variable screening procedure is considered, where all relevant predictors can be consistently identified \citep{Wang2009}.
\citep{zhu2011} developed a unified framework of variable screening for both parametric and semi-parametric models. Variants of SIS are further investigated in the ultra high dimensional settings, such as the factor profile SIS and high-dimensional ordinary least square projection (HOLP) based on Moore-Penrose inverse \citep{wang2012, Wang2016}. Another technique of variable screening is to incorporate robustness into variable screening procedure by introducing robust estimators of correlation in ultra high dimensional settings \citep{li2011, li2012}. A third choice is to introduce tilting procedure into variable screening. \citep{Cho2012} proposed variable screening by tilting procedure based on which the spurious correlation covariates can be effectively controlled, and \citep{Zhao2021dynamic} proposed dynamic tilted current correlation screening by forming a path of covariates entering the model based on the tilted current correlation between the predictor and the current residual. As the best of our knowledge, research gap on detecting change points in ultra high dimensional panel data still remains. 

The other concern lies on how to select penalties in group variables selection that is usually involved in change point detection. Normally, convex penalties such as LASSO are employed in the above group selection procedure. Extensive literature has explored theoretical properties of the group LASSO penalty \citep{buhlmann2011}. In particular, the group LASSO is consistent in group selection for fixed number of covariates under the irresprentable condition or its variant \citep{meinshausen2006, Zhao2006, zou2006}, and the bounds on the prediction and estimation errors are studied under different conditions; see, for example, \citep{bickel2009,wei2010,lounici2011}. While the group LASSO embraces excellent properties regarding prediction, the consistency of its selection lies on the belief that the design matrix are irrespresentable. This assumption, however, may not hold in high-dimensional settings especially when the dimension of covariates is much greater than the sample size, such as in our case. The group LASSO may not achieve selection consistency if the penalty parameter is determined by minimizing the prediction error, and the approach is also likely to select a model which is larger than the true model with relatively high false positive group selection rate \citep{Fan2001,leng2006, huang2012}. Hence, it is reasonable to use non-concave penalties in change point detection. Possible choices are 
the smoothly clipped absolute deviation (SCAD) penalty \citep{Fan2001} and the minimax concave penalty (MCP) \citep{zhang2010}. Both of them enjoy the oracle property during group selection, indicating that the corresponding penalized estimators are equal to the least squares estimator under appropriate conditions with high probability as if the model is known.

In this paper, we propose a two-stage approach to detect change point in ultra high-dimensional setting, by firstly conducting variable screening to reduce the dimension, and then detecting change points in the framework of group variable selection. Not only the spurious correlation between covariates is taken into consideration in variable screening, but non-convex penalties are studied in change point detection in a ultra high dimension. Asymptotic properties are derived, and the numerical investigations show the promising performance of the proposed approach.

The rest of the paper is organized as follows. Sections 2 introduces the methodology of the proposed approach, and in Section 3 the asymptotic properties are derived. Section 4 introduces the implementation of the proposed method, and numerical results are described in Section 5 and 6. We conclude the paper in Section 7.

\section{Methodology}\label{sec:method}

Let $\{(\yit, \xit);~i=1,\cdots, N; t=1, \cdots, T\}$ be a random sample, where $N$ is the total number of subjects, $T$ is the total number of observation time points. For subject $i$ at time $t$, $\yit$ is the response, and $\xit$ is the $p\times 1$ covariate vector with $p\gg N$. The covariates are believed to be linearly related to the response, i.e., we consider a linear model between the response and the covariates
\begin{equation}\label{eq:linear0}
E(\yit | \xit, \alpha_{i})=\alpha_{i}+\xit^{\top} \bfbetat, \quad i=1, \cdots, N ;~t=1, \cdots, T,
\end{equation}
where $E(\yit|\cdot)$ represents the conditional expectation of $\yit$ given $\xit$ and $\alpha_i$, where the intercept $\alpha_i$ is the $i$th individual effect invariant of $t$ and $\bfbetat$ is the unknown $p\times 1$ slope coefficient parameter vector that is to be estimated, possibly changing over time. Without loss of generality, the individual effect represented by $\alpha_i$ can be eliminated after the response and the covariates are standardized, respectively, and \eqref{eq:linear0} shrinks to
\begin{equation}\label{eq:linear}
E(\yit | \xit)=\xit^{\top} \bfbetat, \quad i=1, \cdots, N ;~t=1, \cdots, T.
\end{equation}

To identify change points over time, a popular way is to employ the penalized regression estimator, which is defined as $\hat{\boldsymbol{\beta}}=\arg \min _{\boldsymbol{\beta}}~S_{NT,\lambda}(\boldsymbol{\beta})$, 
where $\boldsymbol{\beta}=(\boldsymbol{\beta}_1,...,\boldsymbol{\beta}_T)$ and 
\begin{equation}\label{eq:S}
S_{N T, \lambda}(\boldsymbol{\beta})= \frac{1}{N T} \sum_{t=1}^{T} (\mathbf{y}_t  - \mathbf{X}_t \boldsymbol{\beta}_t)^{\top}(\mathbf{y}_t - \mathbf{X}_t \boldsymbol{\beta}_t) ~+~ \frac{1}{T}\sum_{t=2}^T p_1(\|\boldsymbol{\beta}_t - \boldsymbol{\beta}_{t-1}\|_2;\lambda_1),
\end{equation}
where $\mathbf{y}_t = (y_{1t},\ldots, y_{Nt})^{\top}$, $\mathbf{X}_t = (\mathbf{x}_{1t}^{\top}, \ldots, \mathbf{x}_{Nt}^{\top})^{\top}$, $p_1(\cdot;\lambda_1)$ 
is a scalar penalty function with  $\lambda_1$ as the tuning parameter, and $\|\mathbf{z}\|_2$ is the $l_2$ norm of a vector $\mathbf{z}$. To solve the optimization problem, however, when the dimension of the input $p$ is much greater than $N$ for a given time point $t$, $(\mathbf{X}_t^{\top}\mathbf{X}_t)^{-1}$ may not exist due to the commonly seen spurious correlations among covariates in time change point detection, even if the penalty term dominates in \eqref{eq:S}; this makes the optimization problem unsolvable. The other concern is the choice of the penalty. Compared with traditional convex penalties in detection of time change points, such as the often employed LASSO penalty, non-convex penalties may be more appropriate to guarantee the asymptotic selection and estimation consistency. In fact, change points may arise when only limited number of covariates tend to change.


Consequently, we propose to adopt a two-stage procedure. In the first stage, we employ a screening technique to reduce the ultra high input dimension to a moderate size, taking spurious correlations among covariates into consideration, and in the second stage, we identify the locations of the change points based on the remaining covariates from the first stage accordingly. We have proved that such a procedure embraces the selection consistency under moderate assumptions and conditions.

\subsection{Variable screening and DTCCS}\label{subsec:screening}
In ultra high-dimensional settings, spurious correlation usually exists between covariates, and marginal metric-based techniques may fail to detect truly relevant covariates such as the SIS and its variants. Thus, during the first stage of the our proposed procedure, we proposed to employ the \textit{dynamic titled current correlation screening} (DTCCS) procedure to screen variables based on the \textit{high-dimensional correlation estimator} (HDCE) whose screening mechanism takes spurious correlation into consideration, possessing the sure screening property and low computational complexity in the ultra high-dimensional setting.

As to employ the DTCCS method and to simultaneously find covariates that are relevant to the response at all time points, we re-organize \eqref{eq:linear} in the following way after centering the response and standardizing the covariates. Denote the stacked response variable $\yit$ by 
$\mathbf{Y}=(\yonedot^{\top},\ldots, \yNdot^{\top})^{\top}$ where $\mathbf{Y}$ is an $NT\times 1$ vector and $\yidot=(y_{i1},\ldots, y_{iT})^\top$, and the corresponding design matrix $\mathbb{X}=(\Xonedot, \ldots, \Xpdot)$ where $\mathbb{X}$ is an $NT\times p$ matrix and $\Xjdot$ is an $NT\times 1$ column vector, $j=1,\ldots,p$. Thus, the linear relation between the response and the covariates is given by
\begin{equation}\label{eq:screen}
\mathbf{Y} = \beta_1 \Xonedot + \cdots + \beta_p \Xpdot + \boldsymbol{\epsilon} = \mathbb{X}\bfbeta + \boldsymbol{\epsilon},
\end{equation}
where $\bfbeta=(\beta_1,\ldots, \beta_p)^\top$ is the $p\times 1$ coefficient vector, and $\boldsymbol{\epsilon} \in \mathbb{R}^{NT}$ is the random noise vector that is usually assumed to follow an i.i.d $N(\mathbf{0}, \sigma^2\mathbf{I}_{NT})$ distribution. Throughout this paper, $\mathbb{X}$ is assumed to have full row rank. Screening variables and keeping relevant covariates in ultra high dimension is equivalent to find the most important nonzero components in $\bfbeta$ in the current stage, while spurious correlation among covariates may lead to inaccurate screening effect \cite{Zhao2021dynamic}. 

Correspondingly, the DTCCS method screens variables by sequentially ranking the importance of variable based on the \textit{high-dimensional correlation estimator} (HDCE), which measures the correlation between the response and each of the left covariates with the spurious correlation tilted by those covariates which are selected in previous steps. 
To be specific, let ${\cal F}=\{1,2,\ldots,p\}$ correspond to the full model with all $p$ covariates, and ${\cal T}=\{j:1\leq j \leq p,\beta_j\neq 0\}$ corresponding to the true model with covariates whose coefficients are truly nonzero where $|{\cal T}|=m_0$. Define $\Mk=\{j_1,\ldots, j_k\}$ as the active set corresponding to a model that contains $\mathbf{X}_{j_1}, \ldots, \mathbf{X}_{j_k}, 1\leq k\leq NT$ as relevant covariates. For the ultra high-dimensional $NT\times p$ design matrix $\mathbb{X}$, to bleach the effect from spurious correlations among covariates, each column $\Xjdot$ is firstly considered as a response variable regressed on the rest $(p-1)$ columns as the corresponding design matrix, denoted as $\Xmj$. Then the $j$-th covariate $\Xjdot$ is tilted by employing the ridge regression such that the effect of other variables $\mathbf{X}_{l\boldsymbol{\cdot}}$ on $\Xjdot$ is much reduced for $l\neq j$. Essentially, this procedure is equivalent to project $\Xjdot$ onto the column space of $\Xmj$ with the projection matrix 
\begin{equation}\label{eq:H}
\mathbb{H}_j = \Xmj(\Xmj^\top \Xmj + \lambda_0 \mathbb{I}_{p-1})^{-1}\Xmj^\top,
\end{equation}
where $\lambda_0$ is a tuning parameter that controls the effect of tilting. Thus, the contribution of each covariate to the response variable $\mathbf{Y}$ eliminating the spurious correlation between covariates is equivalent to that of each variable to the current residual $\mathbf{Z}$ after using ridge regression of $\mathbf{Y}$ against the active variables in the current ${\cal M}_k$ with tuning parameter $\lambda_0'$, and this is measured by the high-dimensional correlation estimator (HDCE), defined as 
\begin{equation}
    \hat{\rho}_j(\lambda_0) = \frac{1}{a_j}\Xmj^\top (\mathbb{I}_{NT} - \mathbb{H}_j)\mathbf{Z},
\end{equation}
where $a_j$ is a normalizing constant that guarantees the tilted correlation bounded by unity. Correspondingly, the variables with absolute HDCEs among the first $[\omega NT]$ largest are selected given a specific $\omega$, where $[\cdot]$ is the integer operator. This shrinks the full model to a submodel ${\cal M}_\omega$ with size $|{\cal M}_\omega| \ll  NT$. 

To finally achieve variable screening, an iterative procedure is employed. Each iteration will return an estimated active set ${\cal M}_k$ for $k=1,\ldots,K$ where $K$ is the number of iterations. Define ${\cal M}_0 = \phi$ and denote ${\cal F}$ as the full model. For $k$-th iteration, the remaining variables in ${\cal F}$ but not in the cumulative active set $\bigcup_{s=1}^{k-1}{\cal M}_s$ are ranked descendingly by their absolute value of HDCEs with $\mathbf{Z}_k$ as the current residual that arises from regressing $\mathbf{Y}$ on the covariates in $\bigcup_{s=1}^{k-1}{\cal M}_s$. The iteration is terminated either when sufficient covariates are selected or the number of iteration attains the upper limit $K$. Such a dynamic procedure is called the DTCCS, and the selected covariates by DTCCS are contained in ${\cal M}=\bigcup_{k=1}^K{\cal M}_k$. Correspondingly, the variable screening is achieved with a moderate size of $m$ active covariates, which will be used for change point detection in the next stage.

\subsection{Change point detection with non-convex penalties}\label{subsec:cpt}
We now describe how to detect change points with the screened variables from the first stage. Suppose ${\cal X}_1,\ldots, {\cal X}_m \in {\cal M}$ are $m$ variables that are selected from the first stage. Based on covariates in ${\cal M}$, we identify the time change points with the following "surrogate" model
\begin{equation}\label{eq:cpt}
   E(\yit | \txit)=\txit^{\top} \bftbetat, \quad i=1, \cdots, N ;~t=1, \cdots, T,
\end{equation}
where $\txit=({\cal X}_{it,1},\ldots,{\cal X}_{it,m} )^\top$, and $\bftbetat$ is the unknown $m\times 1$ parameter vector, possibly changing over time. Stacking the response and the covariates by time and objection, we reorganize \eqref{eq:cpt} into
\begin{equation}\label{eq:stackcpt}
\varY = \varX\bftbeta + \boldsymbol\epsilon,
\end{equation}
where ${\cal Y} = (\mathbf{y}_1^{\top},\ldots, \mathbf{y}_T^{\top})^{\top}$ is the $NT\times 1$ stacked response, 
${\cal X}=diag(\mathbf{X}_1, \ldots, \mathbf{X}_T)$ is the $NT\times Tm$ diagonal block matrix, $\bftbeta=(\tilde{\bm\beta}_1^{\top}, \ldots, \tilde{\bm\beta}_T^{\top})^{\top}$ is the $Tm\times 1$ stacked parameter vector, and $\bm\epsilon = ( \bm\epsilon^{\top}_1, \ldots, \bm\epsilon^{\top}_T)^{\top}$ is the $NT \times 1$ stacked i.i.d. random error with mean  $\mathbf{0}$ and $cov(\bm\epsilon)= diag(\Sigma_1, \ldots, \Sigma_T)$ with $\Sigma_t=cov(\bm\epsilon_t)$ for $t=1,\ldots, T$. Without loss of generality, we assume unity matrix for all $\Sigma_t$. 

To detect change points in time is equivalent to test whether $\bftbetat-\tilde{\bm\beta}_{t-1}=0$ for $t=2,\ldots, T$. Re-parameterizing \eqref{eq:cpt} with $\boldsymbol{\theta}_1=\tilde{\boldsymbol{\beta}}_1$ and $\boldsymbol{\theta}_t=\bftbetat-\tilde{\boldsymbol{\beta}}_{t-1}$ for $t=2,\ldots,T$, $\tilde{\bfbeta}$ can be fully replaced by $\mathbb{A}\boldsymbol{\theta}$, where $\boldsymbol{\theta}=(\boldsymbol{\theta}_1^{\top}, \ldots, \boldsymbol{\theta}_T^{\top})^{\top}$ and $\mathbb{A}$ is a lower-triangular blocked matrix with each block being $m$-dimensional identity matrix $\mathbb{I}_m$. 
Correspondingly, \noindent\eqref{eq:cpt} is reformulated as 
$E(\yit | \txit)=\txit^{\top} \sum_{s=1}^t\boldsymbol{\theta}_s$,
and \eqref{eq:stackcpt} can be reorganized as
\begin{equation}\label{eq:stackcptnew}
\varY = \boldsymbol{\alpha}+ \varX\bftbeta + \boldsymbol\epsilon = {\cal X}^{\star}\boldsymbol{\theta} + \boldsymbol\epsilon,
\end{equation}
where ${\cal X}^{\star}=\varX\mathbb{A}$.
Hence, the change point detection is further equivalent to find nonzero component vector $\boldsymbol{\theta}_j$ in $\boldsymbol\theta$, which appears to be a group selection of non-zero variables in ${\cal M}$.

A popular way of group variable selection is to use penalized regression with an optimization problem as
$\hat{\bm\theta}=\arg \min _{\bm\theta}~S_{NT,\lambda}(\bm\theta)$, where 
\begin{equation}\label{eq:Snew}
S_{N T, \boldsymbol{\lambda}}(\bm\theta) 
= \frac{1}{N T} \left(\varY - {\cal X}^{\star}\boldsymbol{\theta}\right)^\top \left(\varY - {\cal X}^{\star}\boldsymbol{\theta}\right) + \frac{1}{T}\sum_{t=1}^T p_1(\|\boldsymbol{\theta}_t\|_2;\lambda_1).
\end{equation}

Note that in detection of change point, we do not consider the bi-level estimation of coefficients, that is, we only concern about non-zero component vector $\boldsymbol{\theta}_t$ for all $t$ without making sparsity assumption on each element in $\boldsymbol{\theta}_t$. Rather than employing convex penalties for group variable selection in traditional time change point detection, we consider non-convex penalties. In particular, two non-convex forms of penalty are considered, 
namely: (1) the SCAD penalty with $p(x;\lambda,a)=\lambda\int_{0}^{|x|} \min\{1, (a-t/\lambda)_{+}/{(a-1)}\}dt$ for some $a>2$ and (2) the minimax concave penalty (MCP) with $p(x;\lambda,a)=\lambda\int_{0}^{|x|} (1-t/(a\lambda)_{+})dt$, for $a>1$, where $a_{+}$ denotes the positive part for any $a\in\mathbb{R}$. Both the SCAD and MCP penalties enjoy the oracle property for individual variables, indicating that the corresponding penalized estimators are, with a probability tending to 1, equal to the least squares estimator if the model is assumed as known under certain conditions. We propose to apply both of the two penalties for $p_1(\cdot;\lambda_1)$, and we can obtain $l_2$ norm group SCAD and MCP change point detection estimators. With the corresponding estimated $\hat{\bm\theta}$, it is feasible to obtain the estimates of the non-zero coefficients based on which all change points are estimated simultaneously. 

\section{Asymptotic properties}
We now present the theoretical justifications for the proposed procedure. The theorems established will asymptotically guarantee the accuracy of the proposed procedure in the first stage with a sufficiently large sample, and consistently detect the change points. 

\begin{theorem}
    (Accuracy of DTCCS) Suppose that Assumptions A1 and A2 in the appendix hold and there exist positive constants $a_1$, $a_2$ and $C_1$ that are defined in the appendix. Then there exists $\phi_N\in(0,1)$ such that 
    $$
    P\left(\underset{j\in{\cal T}}{\min}|\hat{\rho}_j| \geq \phi_N\right)\geq 1- O\left(m_0\cdot \exp\{-C_1NT\}\right) - m_0\cdot\exp\{-a_1 (NT)^{2(1+\kappa)+\nu-\alpha}\}$$
    and 
    $$P\left({\cal T}\subset {\cal M}\right)=1-m_0 a_2\cdot \exp\{-a_1(NT)^{2(1+\kappa)+\nu-\alpha}\} - O\left(m_0\cdot\exp\{-C_1 NT\}\right).$$
\end{theorem}

\begin{theorem}
    (Asymptotic sure screening) Suppose Condition C1-C4 and Assumption A1 and A2 hold in the appendix in the separated supplementary material. If $2(1+\kappa)>\alpha-\nu$, then the property of asymptotic sure screening is satisfied for DTCCS, i.e.,
    $$
    P\left({\cal T}\subset {\cal M}\right)~\rightarrow ~1~\mbox{as}~N\rightarrow\infty.
    $$
\end{theorem}

\begin{theorem}
    (Screening consistency) Suppose $\log(p-m_0)=o\left(\min\{CNT, a_1(NT)^{2(1+\kappa)+\nu-\alpha}\}\right)$ with a large probability, where $a_1$ and $C$ are defined in Lemma 3.4 and 3.5 in the appendix. Then 
    $$
    P\left(\underset{j\not\in{\cal T}}{\max}|\hat{\rho}_j| \leq \phi_N \leq \underset{j'\in{\cal T}}{\min}|\hat{\rho}_{j'}|\right) ~\rightarrow ~1~\mbox{as}~N\rightarrow\infty,
    $$
    where $\phi_N$ is defined in Theorem 1.
\end{theorem}

Theorem 1 gives the lower bound of accuracy of the HDCE and the accuracy of DTCCS. Theorems 2 and 3 are the asymptotic behavior of the DTCCS. In terms of the second stage, the group variable selection consistency with SCAD and MCP penalties, their asymptotic properties in our time change point detection are quite similar to the well established results. We refer the readers to \citep{Yuan2011} for details. For purpose of completeness, we attach the proof of the group selection consistency in the appendix. 

\section{Implementation}
To implement the proposed procedure, we consider computing methods for the two stages. Overall speaking, LARS can be used to compute variable screening in the first stage, and the group coordinate descent (GCD) can be used to compute the second step. Both of the two methods are standard algorithms. For more details, we recommend readers the papers of \citep{yuan2006} and  \citep{huang2012} for group coordinate descent (GCD). Another issue is the choice of tuning parameters, including the sequence of $\boldsymbol{\lambda}_0=\{\lambda_0^{(1)},\lambda_0^{(2)},\ldots\}$ in the variable screening stage, and $\lambda_1$ in the change point detection stage are to be determined. Specifically for the sequence of tuning parameters $\boldsymbol{\lambda}_0$, the generalized cross-validation is usually adopted, such as in \citep{Fan2001} and \citep{wasserman2009}. To reduce the computational cost, we determine knots selection by extending the idea of using the absolute marginal correlation discussed in \citep{Lockhart2014}, and construct a sequence of order statistics $\lambda_0^{(k)}=f(|\mathbf{X}_j^\top \mathbf{Z}_k|)$ where $f$ is a function that maps the current correlation to $[0,\infty)$ for $k=2,3,\ldots,K$ and set $\lambda_0^{(1)}=\infty$. For $\lambda_1$ and $\lambda_2$ in the second stage, we apply the extended BIC proposed in \citep{chen2008} to determine their values from a pre-specified set in real application. 

\textbf{Remark 1.} In real application, each iteration may pick a fixed number of $d$ covariates, rather than a fixed $\omega$ portion of covariates. A large number of $\omega$ leads to selection of too many covariates into the active set in earlier iterations, which may reduce accuracy, while a small number of $\omega$ leads to that of too few covariates in later iterations, which may reduce selection speed. In this way, after $K$ iterations, the dimension reduces from a large value of $p$ to a moderate value of $m=\min(\sqrt{\frac{p}{NT}}K\log{(NT)}, NT-1)$ \citep{Zhao2021dynamic}.

\textbf{Remark 2.}
To terminate the iteration, we employ the idead of quadratically supported risks (QSR), a unified framework of loss functions for selection consistency proposed in \citep{kim2016}. The QSR includes quadratic loss, Huber loss, quantile loss and logistic loss as speciall cases, by using different sequence of positive number $h_n$ termed as GIC. We suggest that the DTCCS should terminate the model by minimizing the QSR with GIC $h_n=\log{p}/NT$.

\textbf{Remark 3.}
The tuning parameter $\lambda_0$ may be different during iterations. Usually we choose a sequence of $\lambda_0^k$ for $k=1,\ldots,K$ as $\infty = \lambda_0^1\geq\lambda_0^2 \geq \lambda_0^K\geq 0$. A greater $\lambda_0$ will reduce the spurious correlation among covariates more strictly. To see how the turning parameter $\lambda$ controls the spurious correlation among covariates, we refer readers to \citep{Zhao2021dynamic} for details.

\section{Numerical investigation}
In this section, we implement the two-stage proposed method for change point detection, namely, firstly to pick potentially relevant covariates from the original variable space using the DTCCS method for screening, and then to identify the time points by group selection after re-parameterizing the original problem. Note that the key focus is to find the nonzero coefficients in time change point setting, though an accurate estimation of the nonzero components are also achieved with a post selection procedure followed by our proposed method.

\subsection{Synthetic data}
The data are generated from the following models
$y_{it} = \alpha_{i}+\sum_{j=1}^p \sum_{t=1}^T \beta_{t,j}x_{it,j} + u_{it},$ for $i = 1,\cdots, N;~t = 1,\cdots, T,$
where we set the variables and parameters satisfy the following cases
$\alpha_{i}$ are i.i.d. from $N(0 ,1)$, $\textbf{x}_{it}$ i.i.d. from $ N(\textbf{0}_p,\textbf{I}_p)$ and $u_{it}$ i.i.d. from $N(0 ,1)$. The coefficients $\beta_{t,1}$, $\beta_{t,2}$, $\beta_{t,3}$ and $\beta_{t,4}$ are time-varying parameters, where $\beta_{t,1} = \beta_{t,2} =2I\{1\leq t \leq \lfloor T/2\rfloor\}+7I\{ \lfloor T/2 \rfloor+1\leq t\leq T\}$, $\beta_{t,3} = \beta_{t,4} = 2I\{1\leq t \leq \lfloor T/3\rfloor\} +  7I\{ \lfloor T/3 \rfloor+1\leq t\leq 2T/3\} + 2I\{ \lfloor 2T/3 \rfloor+1\leq t\leq T\}$, $\beta_{t,5} = \beta_{t,6} = 5$, $\beta_{t,j} = 0, j = 7,\cdots, p$. Thus, the true change points are $t=[T/2]+1$ and $t=[2T/3]+1$. Intensive simulation studies have been explored, with different combinations of sample size $N$ from $\{20, 40, 80, 200\}$, number of covariates $p$ from $\{20,30,40,80,160,200\}$ and numbers of time points for each subject in the sample $T$ from $\{20,40,80,160\}$. The whole procedure is repeated for 100 times, and the average true detective rate (TDR) and false detective rate (FDR) are partially reported in Tables \ref{tab:simuscreen} and \ref{tab:tfdr2stage}, where the former represents the percentage of repetitions that the all truly useful covariates are selected, and the latter represents the percentage that irrelevant covariates are kept.
A complete record is reported in the appendix. The competitors for the first stage are: (i) SIS with the first $[n/\ln{n}]$ covariates are selected \citep{fanlv2010};(ii) HOLP with with the first $[n/\ln{n}]$ covariates are selected \citep{leng2006};(iii) DTCCS with $m$ as the first $[p/4]$ covariates;  (iv) DTCCS with $m=\min(\sqrt{\frac{p}{NT}}K\log{(NT)}, NT-1)$. The third competitor is set to show an empirical choice of number of covariates to be reserved in the first stage of screening, while the fourth is a theoretical optimal choice.


\begin{table}[h!]
\begin{center}
\begin{tiny}
	\caption{Screening results in the first stage}
	\label{tab:simuscreen} 
	\renewcommand\tabcolsep{8.5pt}
	\begin{tabular}{cccccccccccc}
	\\
	\toprule    
	\multicolumn{4}{c}{ } & 
	\multicolumn{4}{c}{True detective rate} & 	\multicolumn{4}{c}{False detective rate} \\ 
	\cmidrule(r){5-8}
	\cmidrule(r){9-12}
	No. &  p &  T &  N &\romannumeral1   & \romannumeral2 & \romannumeral3 & \romannumeral4 &\romannumeral1   & \romannumeral2 & \romannumeral3 & \romannumeral4 \\
	\midrule

	$1$ & $30$ & $20$ & $20$ & $0.94$ & $0.94$ & $1$ & $1$  & $0.01$ & $0.4$ & $0$ & $0$ \\ 
	$2$ & $30$ & $20$ & $40$ & $1$ & $1$ & $1$ & $1$ & $0.4$ & $0.4$ & $0.4$ & $0.08$ \\ 
	$3$ & $30$ & $40$ & $20$ & $1$ & $1$ & $1$ & $1$ & $0$ & $0.7$ & $0$ & $0$ \\ 
	$4$ & $30$ & $40$ & $40$ & $1$ & $1$ & $1$ & $1$ & $0.4$ & $0.7$ & $0.4$ & $0.28$ \\ 
	$5$ & $40$ & $20$ & $20$ & $0.915$ & $0.915$ & $1$ & $1$  & $0.01$ & $0.4$ & $0$ & $0$ \\ 
	$6$ & $40$ & $20$ & $40$ & $1$ & $1$ & $1$ & $1$ & $0.4$ & $0.4$ & $0.4$ & $0.07$ \\ 
	$7$ & $40$ & $40$ & $20$ & $1$ & $1$ & $1$ & $1$ & $0$ & $0.7$ & $0$ & $0$ \\ 
	$8$ & $40$ & $40$ & $40$ & $1$ & $1$ & $1$ & $1$ & $0.4$ & $0.7$ & $0.4$ & $0.2$ \\ 
	$9$ & $80$ & $20$ & $20$ & $0.875$ & $0.875$ & $1$ & $1$  & $0.02$ & $0.4$ & $0$ & $0$ \\ 
	$10$ & $80$ & $20$ & $40$ & $1$ & $1$ & $1$ & $1$ & $0.4$ & $0.4$ & $0.4$ & $0.02$ \\ 
	$11$ & $80$ & $40$ & $20$ & $1$ & $1$ & $1$ & $1$ & $0$ & $0.7$ & $0$ & $0$ \\ 
	$12$ & $80$ & $40$ & $40$ & $1$ & $1$ & $1$ & $1$ & $0.4$ & $0.7$ & $0.4$ & $0.11$ \\  
	$13$ & $160$ & $20$ & $20$ & $0.75$ & $0.75$ & $1$ & $1$  & $0.04$ & $0.4$ & $0$ & $0$ \\ 
	
	$14$ & $160$ & $20$ & $40$ & $1$ & $1$ & $1$ & $1$ & $0.4$ & $0.4$ & $0.4$ & $0.02$ \\ 
	
	$15$ & $160$ & $40$ & $20$ & $1$ & $1$ & $1$ & $1$ & $0$ & $0.7$ & $0$ & $0$ \\ 
	
	$16$ & $160$ & $40$ & $40$ & $1$ & $1$ & $1$ & $1$ & $0.4$ & $0.7$ & $0.4$ & $0.05$ \\ 
	
	
	
	
	\bottomrule
	\end{tabular}
	\end{tiny}
	\end{center}
\end{table}

\begin{table}[h!]
\begin{tiny}
	\caption{TDR and FDR of change point detection for 3 penalties under 4 screening methods} 
	\label{tab:tfdr2stage}
	\centering
	\renewcommand\tabcolsep{2pt}
	\makebox[\linewidth][c]{
	\begin{tabular}{ccccccccccccccccccccccccccc}
	\\
	\toprule 
	&& 
	\multicolumn{12}{c}{true detective rate} &&
	\multicolumn{12}{c}{false detective rate}  \\
	\cmidrule(r){3-14}
	\cmidrule(r){16-27}
	\multirow{2}{*}{No.} &&
	\multicolumn{4}{c}{LASSO} & 	
	\multicolumn{4}{c}{SCAD}& 
	\multicolumn{4}{c}{MCP} &&
	\multicolumn{4}{c}{LASSO} & 	
	\multicolumn{4}{c}{SCAD}& 
	\multicolumn{4}{c}{MCP} \\ 
	\cmidrule(r){3-6}
	\cmidrule(r){7-10}
	\cmidrule(r){11-14}
	\cmidrule(r){16-19}
	\cmidrule(r){20-23}
	\cmidrule(r){24-27}
	&&\romannumeral1   & \romannumeral2 & \romannumeral3 & \romannumeral4 & \romannumeral1   & \romannumeral2 & \romannumeral3 & \romannumeral4&\romannumeral1   & \romannumeral2 & \romannumeral3 & \romannumeral4 &&
	\romannumeral1   & \romannumeral2 & \romannumeral3 & \romannumeral4 & \romannumeral1   & \romannumeral2 & \romannumeral3 & \romannumeral4&\romannumeral1   & \romannumeral2 & \romannumeral3 & \romannumeral4 \\
	\midrule
	
	$1$ && $0.98$ & $1$ & $1$ & $1$ & $0.98$ & $1$ & $1$ & $1$ & $0.97$ & $0.99$ & $1$ & $1$ &  & $0.30$ & $0.20$ & $0$ & $0$ & $0.10$ & $0.05$ & $0$ & $0$ & $0.10$ & $0.04$ & $0$ & $0$ \\ 
	
    $2$ && $1$ & $1$ & $1$ & $1$ & $1$ & $1$ & $1$ & $1$ & $1$ & $1$ & $1$ & $1$ &  & $0.13$ & $0.08$ & $0$ & $0$ & $0$ & $0$ & $0$ & $0$ & $0$ & $0$ & $0$ & $0$ \\ 
    
    $3$ && $1$ & $1$ & $1$ & $1$ & $1$ & $1$ & $1$ & $1$ & $0.99$ & $1$ &  $1$ & $1$ &  & $0.20$ & $0.10$ & $0$ & $0$ & $0.06$ & $0.01$ & $0$ & $0$ & $0.05$ & $0.02$ & $0$ & $0$ \\ 
    
    $4$ && $1$ & $1$ & $1$ & $1$ & $1$ & $1$ & $1$ & $1$ & $1$ & $1$ & $1$ & $1$ &  & $0.10$ & $0.09$ & $0$ & $0$ & $0.05$ & $0.04$ & $0$ & $0$ & $0$ & $0$ & $0$ & $0$ \\ 
    
    $5$ && $0.94$ & $1$ & $1$ & $1$  & $0.93$ & $0.99$ & $1$ & $1$ & $0.94$ & $0.98$ & $1$ & $1$ &  & $0.21$ & $0.23$ & $0.10$ & $0.06$ & $0.07$ & $0.07$ & $0.07$ & $0.01$ & $0.065$ & $0.02$ & $0.01$ & $0$ \\ 
    
    $6$ &&$1$ & $1$ & $1$ & $1$ & $1$ & $1$ & $1$ & $1$ & $1$ & $1$ & $1$ & $1$ &  & $0.15$ & $0.13$ & $0$ & $0$ & $0.05$ & $0.03$ & $0$ & $0$ & $0.08$ & $0.04$ & $0$ & $0$ \\ 
    
    $7$ && $1$ & $1$ & $1$ & $1$ & $1$ & $1$ & $1$ & $1$ & $0.99$ & $0.99$ & $1$ & $1$ &  & $0.13$ & $0.09$ & $0.05$ & $0.05$ & $0.03$ & $0.03$ & $0.01$ & $0$ & $0.01$ & $0.01$ & $0$ & $0$ \\ 
    
    $8$ && $1$ & $1$ & $1$ & $1$ & $1$ & $1$ & $1$ & $1$ & $1$ & $1$ & $1$ & $1$ &  & $0.08$ & $0.07$ & $0.01$ & $0$ & $0.03$ & $0.02$ & $0.01$ & $0.01$ & $0.03$ & $0.03$ & $0$ & $0.01$ \\ 
    
    $9$ && $0.95$ & $1$ & $1$ & $1$  & $0.95$ & $1$ & $1$ & $1$ & $0.95$ & $1$ & $1$ & $1$ & & $0.31$ & $0.27$ & $0.13$ & $0.05$ & $0.08$ & $0.06$ & $0.06$ & $0.05$ & $0.10$ & $0.07$ & $0.06$ & $0.05$ \\ 
    
    $10$ && $1$ & $1$ & $1$ & $1$ & $1$ & $1$ & $1$ & $1$ & $1$ & $1$ & $1$ & $1$ &  & $0.13$ & $0.10$ & $0.03$ & $0.02$ & $0.05$ & $0.03$ & $0$ & $0$ & $0.04$ & $0.03$ & $0$ & $0$ \\
    
    $11$ && $1$ & $1$ & $1$ & $1$ & $1$ & $1$ & $1$ & $1$ & $1$ & $1$ & $1$ & $1$ &  & $0.15$ & $0.12$ & $0.07$ & $0.07$ & $0.06$ & $0.05$ & $0.01$ & $0$ & $0.06$ & $0.05$ & $0.01$ & $0$ \\ 
    
    $12$ && $1$ & $1$ & $1$ & $1$ & $1$ & $1$ & $1$ & $1$ & $1$ & $1$ & $1$ & $1$ &  & $0.09$ & $0.07$ & $0.03$ & $0.03$ & $0.05$ & $0.03$ & $0.01$ & $0.01$ & $0.05$ & $0.03$ & $0.01$ & $0.01$ \\ 
    
    
    
    
    
    $13$ && $0.96$ & $1$ & $1$ & $1$  & $0.96$ & $1$ & $1$ & $1$ &  $0.99$ & $0.96$ & $1$ & $1$ &  & $0.35$ & $0.28$ & $0.16$ & $0.14$ & $0.18$ & $0.16$ & $0.12$ & $0.10$ & $0.19$ & $0.17$ & $0.10$ & $0.10$ \\ 
    
    $14$ && $1$ & $1$ & $1$ & $1$ & $1$ & $1$ & $1$ & $1$ & $1$ & $1$ & $1$ & $1$ &  & $0.23$ & $0.15$ & $0.09$ & $0.06$ & $0.15$ & $0.10$ & $0.05$ & $0.04$ & $0.15$ & $0.10$ & $0.05$ & $0.04$ \\ 
    
    $15$ && $1$ & $1$ & $1$ & $1$ & $1$ & $1$ & $1$ & $1$ & $1$ & $1$ & $1$ & $1$ &  & $0.20$ & $0.17$ & $0.12$ & $0.11$ & $0.10$ & $0.08$ & $0.05$ & $0.05$ & $0.11$ & $0.07$ & $0.05$ & $0.04$ \\ 
    
    $16$ && $1$ & $1$ & $1$ & $1$ & $1$ & $1$ & $1$ & $1$ & $1$ & $1$ & $1$ & $1$ &  & $0.15$ & $0.12$ & $0.05$ & $0.05$ & $0.11$ & $0.09$ & $0.03$ & $0.02$ & $0.10$ & $0.10$ & $0.03$ & $0.02$ \\ 
    
    
    
    
	\bottomrule
	\end{tabular}}
	\end{tiny}
\end{table}

As is obviously found from the true detective rate, almost all procedures with grouped SCAD, MCP and LASSO perform well in the sense that they tend to correctly find the all true change points under different combination of sample size and observed time points and number of features, after the tuning parameters are selected by the adopted information criteria. When the sample size is large, the performance of all three methods are almost identical, while when the sample size is finite, the grouped LASSO even performs slightly better than the grouped SCAD and MCP, though not significant considering the standard error. 

However, in terms of the false detective rate, the grouped SCAD and MCP methods have shown excellent selection accuracy, especially when the number of variables grows sufficiently large, while in contrast, grouped LASSO tend to pick many some additional irrelevant change points apart from the true ones from the FDR rates, which deteriorates the performance. This agrees with the asymptotic selection accuracy of group selection with the non-convex penalty of which the grouped LASSO may be lack. Besides, the standard deviation of the number of irrelevant change points from the grouped LASSO is much greater than those from the grouped SCAD and MCP. Thus, the grouped SCAD and grouped LASSO procedures are robust in the sense of not over estimating the number of irrelevant time points, which advocates to adoption to these two methods. Additionally we find that as the number of time points increases, the performances of the grouped SCAD and MCP tend to have better performance in the sense of both the number of correctly estimated change points and that of the FDR and the corresponding standard deviations, though performance of all methods deteriorates especially LASSO due to an increasing number of the unknown vectors to be estimated, as expected. 

\subsection{Real data examples}
In this section, we examine our proposed procedure on a real data set that contain the G8 nations' macro economic statistics represented by 149 variables with complete records from 1998 to 2017, including the growth rate and other covariates regarding economy, health, population and development of techniques \cite{kock2013oracle}. We aim to find the statistically significant covariates that are relevant to the economic growth, represented by the (log) annual growth rate of GDP and detect its time change points. The annual growth rates of GDP for the eight nations from 1998 to 2017 are partially illustrated in the form of boxplots in Figure \ref{fig:gdp_box}, and more detailed descriptive statistics are attached in the appendix in the supplementary material. A straight expectation of the change points from the growth rates may be the year 2009 when all nations' GDP growth rates have dropped dramatically, and 2010 when a clear reverse occurs. In other years, the annual growth rates are overall comparable with each other. Thus, 2009 and 2010 seem to be plausible time change points.

We employ our proposed procedure to the data set, where the annual growth rate is set as the response variable and the remaining variables as covariates. The variable screening results are reported in Table \ref{tab:screen}. The employment rate from different economic sectors including agriculture, industry and services, appear to be the most important variables, which agrees with the empirical experience of the global economic growth \cite{kock2013oracle}.
Another group of selected covariates are population-related variables, including labor force participation rate for ages 15-24, population aging from 15 to 64 and that in largest cities. Additionally, the annual growth of the adjusted net total income per capita is selected. A complete list of variables that remain from the first stage of our proposed procedure can be found in the appendix. It is worth noting that the initial GDP has been screened out and is not selected as the relevant feature for later detection of change points, which may not support the neoclassical growth model that a higher initial wealth may lead to a lower growth rate \cite{barro1991economic}. However, \citep{kock2013oracle} pointed out that the hypothesis is more reliable in the sense of explaining the difference of growth rates between developed countries and less developed ones, while in our case, all economies are developed countries. Meanwhile, the initial GDP of each year has been set as the GDP at the end of last year, which may not provide sufficient time period to have the mechanism take effect.

In terms of the time change points, the estimated results from group selection models with different penalties in the second stage appear to be different. With the well tuned $\lambda$s in different penalty functions using BIC, the group lasso only has identified the year 2009 as the single change point, while both the group SCAD and the group MCP have identified the year 2010 as a second change point apart from 2009. An apparent change in 2010 is detected in the boxplot in Table \ref{fig:gdp_box}.

\begin{figure}[h!]   
  \centering
    \includegraphics[scale=0.42]{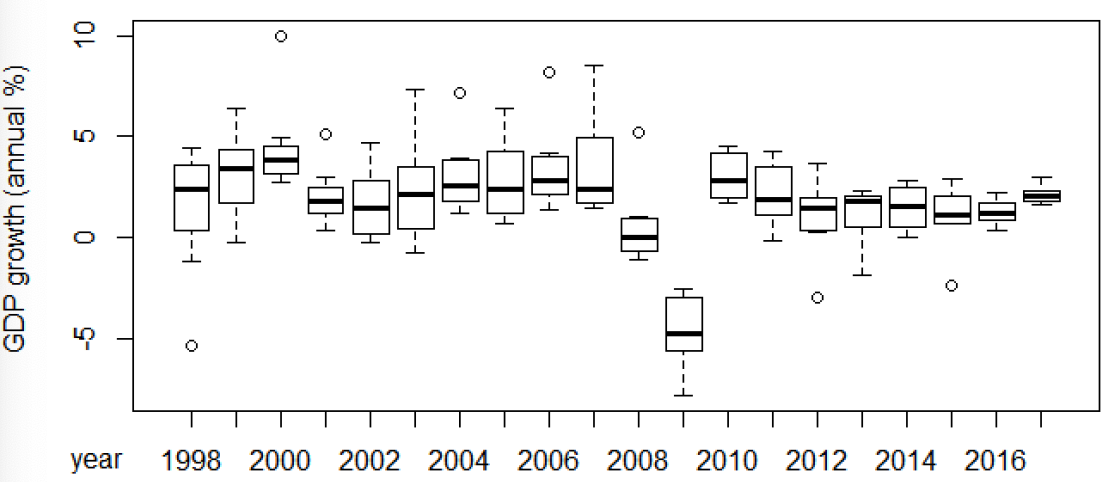}
    \caption{Annual growth rates of G8 nations from 1998 to 2017.} 
    \label{fig:gdp_box}
\end{figure} 


\begin{table}   
  \centering
  \begin{small}
  \caption{Change point detection results with 3 penalties}  \label{tab:screen}
  \begin{tabular}{p{0.7cm}p{3.5cm}p{3.5cm}p{3cm}}
		\toprule
		&  Group LASSO &  Group SCAD &  Group MCP\\
		\midrule
		$1$ & $2009 \quad (BIC)$ &  $2010 $ & $2010$ \\
		$2$ & $2009 \quad 2010$  &  $2009 \quad 2010 \quad (BIC)$ & $2009 \quad 2010 \quad (BIC)$\\
		$3$ &	$2009 \quad 2010 \quad 2012 $& $2008 \quad 2009 \quad 2010$ & $2008 \quad 2009 \quad 2010$ \\
		\bottomrule
	\end{tabular}
	\end{small}
\end{table}

\section{Conclusion}
In this paper, we proposed a two-step procedure for simultaneous detection of change points and estimation of the coefficients in the ultra-high setting. The first step screened the variables from the original ultra-high dimension to a moderately lower one, and the second step determined the change points and estimated the coefficients of the covariates that remain from the screening in the first step. The theoretical properties have been established for the proposed procedure, and the numerical results have supported the superb performance of the procedure.

\newpage

\bibliographystyle{chicago}
\bibliography{ref_dtccs}

\end{document}


\baselineskip  32pt
\doublespacing
\title{Summplementary Materials to Ultra High Dimensional Change Point Detection }

\author{
  Xin~Liu,~Liwen Zhang,~Zhen Zhang\\
  School of Statistics and Management\\
  Shanghai University of Finance and Economics\\
  Shanghai, China, 200433 \\
}

\maketitle

\section*{A. Definitions}
\begin{definition} {(Orthogonal invariance.)}
Let $\mathcal{O}(n)$ be the set of $n\times  n$ orthogonal matrices. An $n$-dimensional random vector $z$ is said to be \emph{orthogonally invariant} if $Qz\overset{(d)}{=}z$ for any orthogonal matrix $Q\in\mathcal{O}(n)$, where the symbol $\overset{(d)}{=}$ represents the equality in distribution. 
\end{definition}

\begin{definition} {(Spherical symmetry.)}
A random vector $z\in\mathbb{R}^{n}$ is said to
be \emph{spherically symmetric} around $\mu\in\mathbb{R}^{n}$ if $z - \mu$ is orthogonally invariant. We denote this as $z\sim\mathbf{ S}_{n}(\mu)$.
\end{definition}

\begin{definition} {(Stiefel manifold.)}
The Stiefel manifold $V_{n}(\mathbb{R}^{p})$ is the set of all orthonormal $n$-frames in an $p$-dimensional Euclidean space. That is $V_{n}(\mathbb{R}^{p})=\{X\in \mathbb{R}^{p\times n}:\, X^{T}X=\mathbf{I}_{n}\}$. The orthogonal group of matrix $\mathcal{O}(n)$ can be considered as a special case of Stiefel manifold which is $V_{n}(\mathbb{R}^{n})$.
\end{definition}

\begin{definition} {(Sub-Gaussian tail condition.)}
Suppose the $\epsilon_{i}$ are independent random variables whose common distribution has a sub-Gaussian tail. That is, there is some $b>0$ such that for every $t\in\mathbb{R}$, we have $E(e^{ t \epsilon_{i}})\leq\exp\{b^{2}t^{2}/2\} $, which implies that there exist positive constants $c_{\epsilon}$ and $d_{\epsilon}$ such that
\vspace{1ex}
\begin{eqnarray}
P\left(\left|\sum_{i=1}^{n}a_{i}\epsilon_{i}\right|>t\right)\leq c_{\epsilon}\cdot \exp\left(-\frac{d_{\epsilon}t^{2}}{\sum_{i=1}^{n}a_{i}^{2}} \label{subG} \right)
\vspace{1ex}
\end{eqnarray}

for all $a=(a_{1},\dots,a_{n})^{T}\in\mathbb{R}^{n}$ and $t>0$.
\end{definition}

\section*{B. Assumptions}
Let $\mathit{vec} (\mathbf{X})$ be the column vector stacked by all the rows of $\mathbf{X}$. Then $\mathit{vec} (\mathbf{X})\sim N(\mathbf{0}, \mathbf{I}_{n}\otimes\Sigma)$ and $\mathit{vec} (\mathbf{B})\sim N(\mathbf{0}, \mathbf{I}_{n}\otimes \mathbf{I}_{p})$, where $\otimes$ denotes the Kronecker product of two matrices. That is, all the elements of $\mathit{vec} (X)$ are standard normal random variables and all the elements of $\mathit{vec} (\mathbf{B})$ are independent and identically distributed standard normal random variables.

\begin{assumption}\label{A1}
 We assume that $var(\mathbf{Y})=O(1)$ and $\lambda=O(p)$ if $\lambda$ is finite, and that the true model size $t_{0}=c_{0} n^{\nu}$ for the sparsity rate $\nu\in[0,1]$.
\end{assumption}

\begin{assumption}\label{A2}{(Concentration Property.)}
 Assume that each entry of the random matrix $\mathbf{B}$ is iid random variables with zero mean and unit variance and that $E|B_{11}|^{4}<\infty$. As $n\rightarrow\infty$, we assume that $p\rightarrow\infty$ and $\frac{n}{p}\rightarrow r\in(0,1)$. Matrix $\mathbf{B}$ is said to have the concentration property if there exist constants $c$, $c_{2} > 1$ and $C_{1} > 0$ such that
 \vspace{1ex}
 \begin{eqnarray}
P\left( d^{\star}(\frac{1}{ \tilde{p}}\tilde{\mathbf{B}}\tilde{\mathbf{B}}^{T})>c_{2}  \text{ and  }   d_{\star}(\frac{1}{ \tilde{p}}\tilde{\mathbf{B}}\tilde{\mathbf{B}}^{T})<1/c_{2} \right)\leq e^{-C_{1}n}\nonumber
\end{eqnarray}
for any $n\times\tilde{p}$ submatrix $\tilde{\mathbf{B}}$ of $\mathbf{B}$ with $cn < \tilde{p}\leq p$.
\end{assumption}

\section*{C. Conditions}
\begin{condition}\label{c1} {(Polynomial high-dimensional)}. $p > n$ and $p = O(n^{\alpha} )$ for some $\alpha > 0$.
 \end{condition}

\begin{condition}\label{c2} {(Normality assumption)}. Assume $\mathbf{X}$ follows the matrix normal distributions and $\epsilon_{i}\sim N(0, \sigma^{2})$ with variance $\sigma^{2}$ for $i=1,\dots, n$.
\end{condition}

\begin{condition}\label{c3} {(Covariance matrix)}. Let $d_{\star}(A)$ and $d^{\star}(A)$ represent the smallest and the largest eigenvalues of the positive definite covariance matrix $A$, respectively. We assume that for some $0\leq\kappa\leq1$ and $c_{1}>0$, the conditional number of $\,\Sigma$, $cond(\Sigma)=d^{\star}(\Sigma)/d_{\star}(\Sigma) \leq c_{1}n^{\kappa}$.
\end{condition}

\begin{condition}\label{c4} {(Tilting parameter)}. Let $\lambda$ be the tilting parameter introduced in Section 2.1, where we require $\lambda=O(p)$ for the finite selection of $\lambda$.
\end{condition}

\section*{D. Proof of Theorems 1 to 3}
\begin{prop}\label{P1} Assume that the conditional number of $\Sigma$, $cond(\Sigma)\leq c_{1}n^{\kappa}$ for some constants $c_{1}$ and $\kappa$ and that $\Sigma_{ii}=1$ for $i=1,2,\dots,p$, then we have
 \begin{eqnarray}
 d_{\star}(\Sigma)\geq c_{1}^{-1} n^{-\kappa} \text{ and  }   d^{\star}(\Sigma)\leq  c_{1} n^{\kappa}.\label{covmatrix}
\end{eqnarray}
\end{prop}

\begin{proof}
Note that $p=tr (\Sigma)=\sum_{i=1}^{p} d_{i}$, so we obtain that $ d_{\star}(\Sigma)\leq1$ and  $d^{\star}(\Sigma)\geq1$. Therefore,
 \begin{eqnarray}
 d_{\star}(\Sigma)\geq \frac{1}{cond(\Sigma)} \text{ and  }   d^{\star}(\Sigma)\leq cond(\Sigma).\nonumber
\end{eqnarray}
\end{proof}

\begin{prop}\label{P2}{(Interlacing inequalities for singular values.)}
Let $\mathbf{B}$ be an $n\times p$ matrix with rank $t=\min(n,p)$, the singular values of $\mathbf{B}$ are the  square roots of the non-zero eigenvalues of the positive semidefinite matrix $\mathbf{B}^{T}\mathbf{B}$ (or $\mathbf{B}\mathbf{B}^{T})$. The sequence of singular values is $\sigma_{1}(\mathbf{B})\geq\dots\geq\sigma_{t}(\mathbf{B})>0=\dots=0$. The relationship between any $(n-s)\times (p-r)$ submatrix $\tilde{\mathbf{B}}$ and $\mathbf{B}$ is
 \begin{eqnarray}
  \sigma_{k}(\mathbf{B})\geq \sigma_{k}(\tilde{\mathbf{B}})\geq \sigma_{k+r+s}(\mathbf{B}) \text{ for integer  }    1\leq k\leq t. \nonumber
\end{eqnarray}
\end{prop}

\begin{lemma}\label{lemma1} 
Let $\mathcal{O}(n)$ be the set of $n\times  n$ orthogonal matrices. A singular value decomposition of the $n\times p$ full rank matrix $\mathbf{B}$ can be expressed as $\mathbf{B}=U D V$, where $U\in\mathcal{O}(n)$, $V\in\mathcal{O}(p)$, and $D=[D_{ij}]$ is an $n\times p$ matrix with  $D_{ij}=0$ for $i\neq j$ and $D_{11}\geq D_{22}\geq\dots\geq D_{nn}$. Let $b_{i}^{T}$ denote the $ith$ row of $\mathbf{B}$ for $i=1,2,\dots, n$. We assume that the $b_{i}^{T}$ are independent and orthogonally invariant, then the distribution of $\mathbf{B}$ is also invariant under $\mathcal{O}(p)$, i.e., $\mathbf{B}Q\overset{(d)}{=}\mathbf{B}$ for any $Q\in \mathcal{O}(p)$.
\end{lemma}

\begin{lemma}\label{lemma2} For any $n\times\tilde{p}$ submatrix $\tilde{\mathbf{X}}$ of $\mathbf{X}$ with $n < \tilde{p}\leq p$,
non-zero eigenvalues of  $\,\tilde{\mathbf{X}}\tilde{\mathbf{X}}^{T}$ are bounded by $c_{1}^{-1}n^{-\kappa}  d_{\star}(\tilde{\mathbf{B}}\tilde{\mathbf{B}}^{T} )$ and $c_{1}n^{\kappa} d^{\star}(\tilde{\mathbf{B}}\tilde{\mathbf{B}}^{T} )$ where $d_{\star}(A)$ and $d^{\star}(A)$ are the smallest and the largest eigenvalues of the positive definite matrix $A$ and  $c_{1}$ is a positive constant.
\end{lemma}

\begin{lemma}\label{lemma3} Suppose Assumptions \ref{A1} and \ref{A2} hold. Then for $C>0$ and for any unit norm vector $v$, there exist constants $c_{3}$ and $c_{4}$ with $0<c_{3}<1<c_{4}$ such that
 \begin{eqnarray}
P\left(| e_{j}^{T}\Xi_{j} v| <c_{3} n^{1-\kappa/2} \text{ or  }  |e_{j}^{T}\Xi_{j} v| >c_{4} n^{1+\kappa/2} \right)\leq 4 e^{-C n}, \nonumber
\end{eqnarray}
where $\Xi_{j}=\mathbf{X}^{T} \mathbf{P}_{j}\mathbf{F}_{j}\mathbf{P}_{j}^{T}\mathbf{X}$.
\end{lemma}

\begin{lemma}\label{lemma4} Suppose Assumptions \ref{A1} and \ref{A2} hold. Then for any $C>0$, there exists constants $c_{5}$ and $ c_{6}>0$ such that for any $j=1,2,\dots,p$,
\begin{eqnarray}
P\left( | e_{j}^{T}\Xi_{j} \beta| < c_{5} n^{1-\kappa} \right)\leq O(e^{-C n}), \nonumber
\end{eqnarray}
and
\begin{eqnarray}
P\left(  |e_{j}^{T}\Xi_{j} \beta |> c_{6} n^{1+\kappa+\nu/2} \right)\leq O(e^{-C n}). \nonumber
\end{eqnarray}
\end{lemma}

\begin{lemma}\label{lemma5}
Suppose Condition \ref{c2}, and Assumptions \ref{A1} and \ref{A2} hold and let $\nu$ be the sparsity rate, i.e., $t_{0}=c_{0} n^{\nu}$ is the true model size. Assume $2(1+\kappa) >\alpha-\nu$ where $\alpha$ is defined in Condition \ref{c1}. Then for any $C>0$, there exists some constants $c_{8}$ and $ c_{8}^{\prime}>0$ such that for any $j=1,2,\dots,p$,
\vspace{1ex}
\begin{eqnarray}
P\left(  |\eta_{j}| > c_{8} n^{1+\kappa+\nu/2} \right)\leq c_{\epsilon} \exp\{ - c_{8}^{\prime} n^{2(1+\kappa)+\nu-\alpha} \}. \nonumber
\end{eqnarray}
\vspace{1ex}
Also, there exists some small positive $c_{9}$, $c_{9}=o(n^{\kappa-1})$, 
such that for any $j=1,2,\dots,p$,
\begin{eqnarray}
P\left(  |\eta_{j}| < c_{9} n^{1-\kappa} \right)\leq  O(e^{-C n}). \nonumber
\end{eqnarray}
\end{lemma}
\vspace{1ex}

\begin{prop}\label{P3}
Let $\chi^{2}_{1},\dots,\chi^{2}_{n}$ be iid $\chi^{2}$-distributed random variables. Then,

(i) for any $\epsilon>0$, we have
 \begin{eqnarray}
P\left(\frac{1}{n} \sum_{i=1}^{n}\chi_{2}^{2}>1+\epsilon\right)\leq e^{-A_{\epsilon}n}, \nonumber
\end{eqnarray}
where $A_{\epsilon}=[\epsilon- \log(1+\epsilon)]/2>0$.

(ii) for any $\epsilon>0$, we have
 \begin{eqnarray}
P\left(\frac{1}{n} \sum_{i=1}^{n}\chi_{2}^{2}< 1-\epsilon\right)\leq e^{-B_{\epsilon}n}, \nonumber
\end{eqnarray}
where $B_{\epsilon}=[-\epsilon- \log(1-\epsilon)]/2>0$.
\end{prop}

\begin{prop}\label{P4}
Let $e_{i}=(0,\dots,0,1,0,\dots,0)^{T}$ denote the $ith$ natural base in the $p$ dimensional space.
Let $V$ be uniformly distributed on $\mathcal{O}(p)$ and $\tilde{V}^{T}$ be the top $n$ rows of $V$ with $\tilde{V}\in V_{n}(\mathbb{R}^{p})$. Then for any $C>0$, there exist $c_{1}^{\prime}$, $c_{2}^{\prime}$ with $0<c_{1}^{\prime}<1<c_{2}^{\prime} $, such that
\vspace{1ex}
 \begin{eqnarray}
P\left(e_{1}^{T}\tilde{V}\tilde{V}^{T}e_{1}<c_{1}^{\prime}\frac{n}{p} \text{ or  }  e_{1}^{T}\tilde{V}\tilde{V}^{T}e_{1}>c_{2}^{\prime}\frac{n}{p} \right)\leq 4 e^{-C n}. \nonumber
\end{eqnarray}
\end{prop}

\begin{proof}
$\tilde{V}=V^{T}
\left(
\begin{array}{c}
I_{n} \\
\textbf{0}_{p-n,n}
\end{array}
\right)_{p\times n}$, and its transpose $\tilde{V}^{T}=\left(
\begin{array}{c:c}
I_{n}&\textbf{0}_{n,p-n} 
\end{array}
\right)_{n\times p} V$.

$V e_{1}$ is the first column of $V$ and is uniformly distributed on a unit sphere. Let $\omega_{i}, i=1,2,\dots,p$, be independent and identically distributed random variable from standard normal distribution, we have
\begin{eqnarray}
V e_{1} \overset{(d)}{=} \left(\sqrt{\sum_{j=1}^{p}\omega_{j}^{2}}\right)^{-1/2} \left(\omega_{1}, \omega_{2}, \dots, \omega_{p} \right)^{T}\nonumber
\end{eqnarray}
 and
 \begin{eqnarray}
 \tilde{V}^{T}e_{1} \overset{(d)}{=} \left(\sqrt{\sum_{j=1}^{p}\omega_{j}^{2}}\right)^{-1/2} \left(\omega_{1}, \omega_{2}, \dots, \omega_{n} \right)^{T}.\nonumber
 \end{eqnarray}
Hence $e_{1}\tilde{V} \tilde{V}^{T}e_{1} \overset{(d)}{=} \frac{\omega_{1}^{2}+\dots+\omega_{n}^{2}}{\omega_{1}^{2}+\dots+\omega_{p}^{2}}$, which is a random variable following a beta distribution with parameter $n/2$ and $(p-n)/2$. By Proposition \ref{P3}, we know that for any $C>0$, there exists some $\epsilon>0$ such that
\vspace{1ex}
\begin{eqnarray}
P\left(\frac{1}{n} \sum_{i=1}^{n}\omega_{i}^{2}< 1-\epsilon\right)\leq e^{-Cn}, \text{   } P\left(\frac{1}{p} \sum_{i=1}^{p}\omega_{i}^{2}> 1+\epsilon\right)\leq e^{-Cp}<e^{-Cn}, \nonumber
\end{eqnarray}
and
 \begin{eqnarray}
P\left(\frac{1}{n} \sum_{i=1}^{n}\omega_{i}^{2}> 1+\epsilon\right)\leq e^{-Cn}, \text{   } P\left(\frac{1}{p} \sum_{i=1}^{p}\omega_{i}^{2}< 1-\epsilon\right)\leq e^{-Cp}<e^{-Cn}. \nonumber
\end{eqnarray}
Let $c_{1}^{\prime}=\frac{1-\epsilon}{1+\epsilon}$ and $c_{2}^{\prime}=\frac{1+\epsilon}{1-\epsilon}$. By Bonferroni's inequality, we have
 \begin{eqnarray}
P\left(e_{1}^{T}\tilde{V}\tilde{V}^{T}e_{1}<c_{1}^{\prime}\frac{n}{p} \text{ or  }  e_{1}^{T}\tilde{V}\tilde{V}^{T}e_{1}>c_{2}^{\prime}\frac{n}{p} \right)\leq 4 e^{-C n}. \nonumber
\end{eqnarray}
\end{proof}

\begin{proof}[\textbf{Proof of Lemma \ref{lemma2}:}]
Let $\tilde\Sigma$ be a $\tilde{p}\times\tilde{p}$ submatrix of $\Sigma$. By Propositions \ref{A1} and \ref{A2},
$$c_{1}^{-1}n^{-\kappa}  d_{\star}(\tilde{\mathbf{B}}\tilde{\mathbf{B}}^{T} ) I_{n}\leq \,\tilde{\mathbf{X}}\tilde{\mathbf{X}}^{T}=\tilde{\mathbf{B}}\tilde\Sigma\tilde{\mathbf{B}}^{T}\leq  c_{1}n^{\kappa} d^{\star}(\tilde{\mathbf{B}}\tilde{\mathbf{B}}^{T} ) I_{n}.$$
\end{proof}

\begin{proof}[\textbf{Proof of Lemma \ref{lemma3}:}]

Recall $\Xi_{j}=\mathbf{X}^{T} \mathbf{P}_{j}\mathbf{F}_{j}\mathbf{P}_{j}^{T}\mathbf{X}$, $\mathbf{X}=\mathbf{B}\Sigma^{1/2}$ and $\mathbf{F}_{j}=\mbox{diag}(\frac{\lambda}{\lambda+d_{j1}},\ldots,\frac{\lambda}{\lambda+d_{jn}}).$
There exists two $q\times q$ orthogonal matrices $Q_{1}$ and $Q_{2}$ that respectively rotate $\Sigma^{1/2}e_{j}$ and $\Sigma^{1/2}v$ to the same direction of $e_{1}$, i.e., $\Sigma^{1/2}e_{j}=\|\Sigma^{1/2}e_{j}\|_{2}Q_{1}e_{1}$ and $\Sigma^{1/2}v=\|\Sigma^{1/2}v\|_{2}Q_{2}e_{1}$. Then we have
\begin{eqnarray}
 |e_{j}^{T}\Xi_{j}v|=\|\Sigma^{1/2}v\|\cdot\|\Sigma^{1/2}e_{j}\|\cdot e_{1}^{T}Q_{1}^{T}\mathbf{B}^{T} \mathbf{P}_{j}\mathbf{F}_{j}\mathbf{P}_{j}^{T} \mathbf{B}Q_{2} e_{1}.\label{rotation}
\end{eqnarray}

By Lemma \ref{lemma1}, rows of $\mathbf{B}$ are independent and orthogonally invariant, then the distribution of $\mathbf{B}$ is also invariant under $\mathcal{O}(p)$, i.e., $\mathbf{B}Q\overset{(d)}{=}\mathbf{B}$ for any $Q\in \mathcal{O}(p)$.
Rewrite $\mathbf{B}=U\, diag(D_{11},\dots,D_{nn})\tilde{V}^{T}$, where $\tilde{V}^{T}=(I_{n},\textbf{0}_{n,p-n})_{n\times p} V$ and $\tilde{V}\in V_{n}(R^{p})$. It is obvious that $p\cdot[d_{\star}(\frac{1}{p}\mathbf{B}\mathbf{B}^{T} )]I_{n}\leq diag(D_{11}^{2},\dots,D_{nn}^{2}) \leq p\cdot[d^{\star}(\frac{1}{p}\mathbf{B}\mathbf{B}^{T} )]I_{n}$. For the norm of the vector $\|\Sigma^{1/2}v\|$ and $\|\Sigma^{1/2}e_{j}\|$, we have
 \begin{eqnarray}
d_{\star}(\Sigma)\leq v^{T}\Sigma v=\|\Sigma^{1/2}v\|^{2}\leq d^{\star}(\Sigma) \text{    and    }  e_{j}^{T}\Sigma e_{j}=\|\Sigma^{1/2}e_{j}\|^{2}=1.   \nonumber
\end{eqnarray}
By Assumptions \ref{A1} and \ref{A2}, Proposition \ref{P2} and Lemma \ref{lemma2}, the diagonal of $\mathbf{F}_{j}$ is bounded by $O(1)$ and $1$. Since $c_{1}^{-1} n^{-\kappa} \leq d_{\star}(\Sigma)\leq d^{\star}(\Sigma)\leq c_{1} n^{\kappa}$ and $P\left( d^{\star}(\frac{1}{ \tilde{p}}\tilde{\mathbf{B}}\tilde{\mathbf{B}}^{T})>c_{2}  \text{ and  }   d_{\star}(\frac{1}{ \tilde{p}}\tilde{\mathbf{B}}\tilde{\mathbf{B}}^{T})<1/c_{2} \right)\leq e^{-C_{1}n}$ by Proposition \ref{P1} and Assumption \ref{A2}, along with Proposition \ref{P4} and Bonferroni's inequality, we get
 \begin{eqnarray}
P\left( |e_{j}^{T}\Xi_{j} v| <c_{3} n^{1-\kappa/2} \text{ or  }  |e_{j}^{T}\Xi_{j} v| >c_{4} n^{1+\kappa/2} \right)\leq 4 e^{-C n}. \nonumber
\end{eqnarray}

\end{proof}

\begin{proof}[\textbf{Proof of Lemma \ref{lemma4}:}]

Let $\beta=\frac{\beta}{\|\beta\|_{2}}\cdot\|\beta\|_{2}$. Apply the result of Lemma \ref{lemma3}, for $v=\frac{\beta}{\|\beta\|_{2}}$,
 \begin{eqnarray}
P\left( |e_{j}^{T}\Xi_{j} \frac{\beta}{\|\beta\|_{2}}| <c_{3} n^{1-\kappa/2} \right)\leq O( e^{-C n}). \nonumber
\end{eqnarray}
With probability at least $1- O(e^{-Cn})$, $ |e_{j}^{T}\Xi_{j} \frac{\beta}{\|\beta\|_{2}}|\geq c_{3} n^{1-\kappa/2}$. By Assumption \ref{A1} and Proposition \ref{P1}, $var(\mathbf{Y})=O(1)$ and $d^{\star}(\Sigma)\leq c_{1} n^{\kappa}$. 
Then we have $c_{1} n^{\kappa} \| \beta\|_{2}^{2}\geq \|\beta\|_{2}^{2}\, d^{\star}(\Sigma)\geq \beta^{T}\Sigma\beta=var(Y)-\sigma^{2}\geq c_{7}$ for some constant $c_{7}$. Therefore, with probability at least $1- O(e^{-Cn})$, $\|\beta\|_{2} \geq c_{7}^{\prime} n^{-\kappa/2}$ and $e_{j}^{T}\Xi_{j} \beta\geq c_{5} n^{1-\kappa}$ for some constants $c_{7}^{\prime}$ and $c_{5}$ respectively. 
Hence, $P\left( \left|e_{j}^{T}\Xi_{j} \beta\right| < c_{5} n^{1-\kappa} \right)\leq O(e^{-C n})$. Apply the result of Lemma \ref{lemma3}, for $v=e_{i}$, $i=1,2,\dots, p$,
 \begin{eqnarray}
P\left( |e_{j}^{T}\Xi_{j} e_{i}| <c_{3} n^{1-\kappa/2} \text{ or  }  |e_{j}^{T}\Xi_{j} e_{i}| >c_{4} n^{1+\kappa/2} \right)\leq 4 e^{-C n}. \nonumber
\end{eqnarray}

We know the sparsity rate $s=c_{0} n^{\nu}$ from Assumption \ref{A1},
 \begin{eqnarray}
 |e_{j}^{T}\Xi_{j} \beta| & =& \left| \sum_{i\in\mathcal{T}}e_{j}^{T}\Xi_{j}e_{i} \beta_{i} \right|\leq   \sum_{i\in\mathcal{T}}\{ |e_{j}^{T}\Xi_{j}e_{i}|\cdot |\beta_{i}|\}   \nonumber \vspace{1ex} \\ 
 &\leq& \sqrt{\sum_{i\in\mathcal{T}} |e_{j}^{T}\Xi_{j}e_{i}|^{2} }\cdot\|\beta\|_{2}\leq c_{6}n^{1+\kappa+\nu/2},
 \end{eqnarray}
 with probability at least $1- O(e^{-Cn})$. Hence, $P\left( | e_{j}^{T}\Xi_{j} \beta| > c_{6} n^{1+\kappa+\nu/2} \right)\leq O(e^{-C n})$.

\end{proof}

\begin{proof}[\textbf{Proof of Lemma \ref{lemma5}:}]

Define $\eta_{j}= a \epsilon$, where $a=e_{j}^{T}\mathbf{X}^{T} \mathbf{P}_{j}\mathbf{F}_{j}\mathbf{P}_{j}^{T}$ and $\epsilon_{i}\sim N(0,\sigma^{2})$ for $i=1,2,\dots,n$ where $\epsilon_{i}$ is the ith element of $\epsilon$. For the norm square of $a$, we have, with probability at least $1- O(e^{-C_{1}n})$,
\begin{eqnarray}
\| e_{j}^{T}\mathbf{X}^{T} \mathbf{P}_{j}\mathbf{F}_{j}\mathbf{P}_{j}^{T}\|_{2}^{2}&=& e_{j}^{T}\Sigma^{1/2}\mathbf{B}^{T}\mathbf{P}_{j}\mathbf{F}_{j}\mathbf{P}_{j}^{T}\mathbf{P}_{j}\mathbf{F}_{j}\mathbf{P}_{j}^{T}\mathbf{B}  \Sigma^{1/2}e_{j} \nonumber\\ &\leq& p\cdot\|\Sigma^{1/2}e_{j}\|_{2}^{2}\cdot[d^{\star}(\frac{1}{p}\mathbf{B}\mathbf{B}^{T} )]  \nonumber\\
&=& p\cdot [d^{\star}(\frac{1}{p}\mathbf{B}\mathbf{B}^{T} )]. \label{epart1}
   \end{eqnarray}
By Assumption \ref{A2}, the lower bound of $\| e_{j}^{T}\mathbf{X}^{T} \mathbf{P}_{j}\mathbf{F}_{j}\mathbf{P}_{j}^{T}\|_{2}^{2}$ is of the same order as the upper bound. According to the sub-Gaussian tail assumption, we have $P\left(|a\epsilon|>t\right)\leq c_{\epsilon}\cdot \exp\left(-\frac{d_{\epsilon}t^{2}}{\|a\|^{2}}\right)$.
 By choosing $t=c_{8}  n^{1+\kappa+\nu/2}$,
\begin{eqnarray}
P\left(  |a\epsilon| > c_{8} n^{1+\kappa+\nu/2} \right)\leq c_{\epsilon} \exp\{ - c_{8}^{\prime} n^{2(1+\kappa)+\nu-\alpha} \}, \label{epart2}
\end{eqnarray}
where $c_{8}^{\prime}\propto c_{8}^{2}d_{\epsilon}/c_{2}$. Since $\epsilon_{i}\sim N(0,\sigma^{2})$, then $a\epsilon=\sum_{i=1}^{n}a_{i}\epsilon_{i}\sim N(0,\|a\|^{2}\sigma^{2})$. By the property of Normal distributions, $P\left(|a\epsilon|<c_{9} n^{1-\kappa}\right)\rightarrow 0$ as $c_{9} n^{1-\kappa}\rightarrow 0$. Thus,
\begin{eqnarray}
P\left(  |\eta_{j}| < c_{9} n^{1-\kappa} \right)\leq  O(e^{-C n}). \nonumber
\end{eqnarray}

\end{proof}

\begin{proof}[\textbf{Proof of Theorem 1:}]
Applying Lemmas \ref{lemma4} and \ref{lemma5} to all $j\in\mathcal{T}$, for $t_{0}=c_{0}n^{\nu}$, we have
\begin{eqnarray}
P\left(\displaystyle{\min_{j\in\mathcal{T}}} |\xi_{j}| < c_{5}^{\prime} n^{1-\kappa} \right)= O(t_{0}\cdot\exp(-C_{1} n)), \nonumber
\end{eqnarray}
and
\begin{eqnarray}
P\left( \displaystyle{\max_{j\in\mathcal{T}}} |\eta_{j}| > c_{8} n^{1+\kappa+\nu/2} \right)\leq t_{0}c_{\epsilon}\cdot\exp\{ - c_{8}^{\prime} n^{2(1+\kappa)+\nu-\alpha} \}. \nonumber
\end{eqnarray}
If we choose $\theta_{n}\in(c_{0}^{\prime}n^{2\kappa+\nu/2},1)$, which is a rate between $\frac{c_{8} n^{1+\kappa+\nu/2}}{c_{5}^{\prime} n^{1-\kappa}}$ and $1$, then we have
\begin{eqnarray}
P\left( \displaystyle{\min_{j\in\mathcal{T}}} |\hat\rho_{j}|<\theta_{n} \right) &=&P\left( \displaystyle{\min_{j\in\mathcal{T}}} |\xi_{j}+\eta_{j}|<a_{j}\theta_{n} \right)\nonumber\\
&\leq&P\left(\displaystyle{\min_{j\in\mathcal{T}}} |\xi_{j}| < c_{5}^{\prime\prime} n^{1-\kappa} \right)+P\left( \displaystyle{\max_{j\in\mathcal{T}}} |\eta_{j}| > c_{8}^{\prime\prime} n^{1+\kappa+\nu/2} \right) \nonumber\\
&\leq&  O(t_{0}\cdot\exp(-C_{1} n))+t_{0} c_{\epsilon}\cdot\exp\{ - c_{8}^{\prime} n^{2(1+\kappa)+\nu-\alpha} \}. \nonumber
\end{eqnarray}

The first inequality holds since $a_{j}$ is on the same order of $|\xi_{j}|$ for $\lambda\neq0$ and the fact that if one event implies another, it has a smaller probability. The second inequality follows from Lemma 3.4 and 3.5. The detail of the first inequality is 
\begin{eqnarray}
\displaystyle{\min_{j\in\mathcal{T}}} |\xi_{j}+\eta_{j}|<a_{j}\theta_{n} &\Rightarrow &\displaystyle{\min_{j\in\mathcal{T}}} |\xi_{j}|- \displaystyle{\min_{j\in\mathcal{T}}}|\eta_{j}|<a_{j}\theta_{n} \nonumber\\
&\Rightarrow & \displaystyle{\min_{j\in\mathcal{T}}} |\xi_{j}|- \displaystyle{\max_{j\in\mathcal{T}}}|\eta_{j}|<a_{j}\theta_{n}\nonumber\\
&\Rightarrow& \displaystyle{\min_{j\in\mathcal{T}}} |\xi_{j}|< M \text{ or } \displaystyle{\max_{j\in\mathcal{T}}}|\eta_{j}|>m, \nonumber
\end{eqnarray}

where $M$, $m$ can reach $c_{5}^{\prime\prime} n^{1-\kappa}$  and $c_{8}^{\prime\prime} n^{1+\kappa+\nu/2}$ respectively.

Hence,
\begin{eqnarray}
P\left( \mathcal{T}\subset\mathcal{M}_{\theta_{n}}\right)=1-  t_{0}c_{\epsilon}\cdot\exp\{ - c_{8}^{\prime} n^{2(1+\kappa)+\nu-\alpha} \}-O(t_{0}\cdot\exp(-C_{1} n)). \nonumber
\end{eqnarray}
\end{proof}

\begin{proof}[\textbf{Proof of Theorem 2:}]
Apply Theorem 1, $t_{0}\cdot\exp(-C_{1} n)\leq (c_{0}n)/(e^{C_{1}n}) \rightarrow0$ as $n\rightarrow\infty$. Since $2(1+\kappa) >\alpha-\nu$, we have  $\exp\{- c_{8}^{\prime} n^{2(1+\kappa)+\nu-\alpha}\}\rightarrow0$ as $n\rightarrow\infty$. This completes the proof of Theorem 2.
\end{proof}

\begin{proof}[\textbf{Proof of Theorem 3:}]
The same as Theorem 1,
\begin{eqnarray}
P\left(\displaystyle{\max_{j\notin\mathcal{T}}} |\xi_{j}| > c_{6}^{\prime} n^{1+\kappa+\nu/2} \right)\leq (p-t_{0})\cdot O(e^{-C n}), \nonumber
\end{eqnarray}
and
\vspace{1ex}
\begin{eqnarray}
P\left( \displaystyle{\max_{j\notin\mathcal{T}}} |\eta_{j}| > c_{8}^{\prime} n^{1+\kappa+\nu/2} \right)\leq (p-t_{0})\cdot\exp\{- c_{8}^{\prime} n^{2(1+\kappa)+\nu-\alpha} \}+ (p-t_{0})\cdot\exp(-C_{1} n). \nonumber
\end{eqnarray}

Now, if $\theta_{n}$ is chosen as the same as in the Theorem 1, then
\vspace{1ex} 
\begin{eqnarray}
P\left( \displaystyle{\min_{j\notin\mathcal{T}}} |\hat\rho_{j}|> \theta_{n} \right) \leq (p-t_{0})\cdot\exp\{- c_{8}^{\prime} n^{2(1+\kappa)+\nu-\alpha} \}+ O((p-t_{0})\cdot\exp(-C_{1} n)). \nonumber
\end{eqnarray}
\vspace{1ex}
Then, for $\log(p-t_{0})=o(\min\{Cn,  c_{8}^{\prime} n^{2(1+\kappa)+\nu-\alpha} \})$  and combining with Theorem 1, we have
\begin{eqnarray}
P( \displaystyle{\min_{j\in\mathcal{T}}} |\hat\rho_{j}|\geq\theta_{n}\geq \displaystyle{\max_{j\notin\mathcal{T}}} |\hat\rho_{j}| )\rightarrow 1\text{ as } n\rightarrow\infty.\label{larsvs}
\end{eqnarray}
\end{proof}

\section*{E. Theorems of group variable selection consistency}
We now provide theoretical justification of change point detection with 2-norm group MCP, and the case for the proposed procedure with the SCAD penalty can be justified similarly. Define $\hat{\boldsymbol{\theta}}_{MCP}$ as the global minimizer of (9) in the conference paper, with $p_1$ as the MCP penalty. Let ${\cal X}=\left(\varX_1, \ldots,
\varX_T\right)$, and $\Sigma=\varX^\top\varX/NT$. For any $A\subset \{1,\ldots, T\}$, define 
$\varX_A=(\varX_j,j\in A)$ and $\Sigma_A=\varX_A^\top \varX_A /NT$. Denote the true value of $\boldsymbol{\theta}$ as 
$\bftheta^0 = ({\bftheta_1^0}^\top, \ldots, {\bftheta_T^0}^\top)^\top$. Define the set of change point with nonzero coefficients as $S=\{t: \|\bftheta_t^0\|_2\neq 0,~2\leq t\leq T\}$. Note that the first time point is not included in the change point set even if $\|\bftheta_1^0 \|_2\neq 0$. Let $\theta_{\star}^0=\min\{\|\bftheta_t^0\|_2^2/ \sqrt{m_0}:t\in S\}$, and $\bftheta_{\star}^0=\infty$ if $S$ is an empty set. Let $d_{\min}$ be the smallest eigenvalue of $\Sigma$ and set $d_1$ and $d_2$ as the smallest and largest eigenvalues of $\Sigma_S$, respectively. For any $A\subset \{1,\ldots,T\}$, let $g_{\min}(A) = t$, and $g_{\min}(A) = \infty$ if $A$ is empty. Define the function 
\begin{equation}\label{eq:hfun}
    h(u,v)=\exp\left(-v(\sqrt{2u-1}-1)^2/4)\right),~u>1, v=1,2,\ldots,
\end{equation}
which arises from an upper bound for the tail probabilities of chi-square distributions given in Lemma 6 in the Appendix, based on exponential inequality for chi-square random variable of \citet{laurent2000}. Let 
\vspace{1ex} 
\begin{equation}\label{eq:psi1}
    \psi_{1}(\lambda) = \left(T-|S|\right)\cdot h\left(\lambda^2NT/\sigma^2, g_{\min}(S^C)\right)
\end{equation}
\vspace{1ex} 
and 
\vspace{1ex} 
\begin{equation}\label{eq:psi2}
    \psi_{2}(\lambda) =
    |S|\cdot h\left(c^{\star}NT(\theta_{\star}^0-a\lambda)^2/\sigma^2, g_{\min}(S)\right).
\end{equation}
\vspace{1ex} 
Then we have the following result.
\begin{prop}\label{P5}
    Suppose $\epsilon_1, \epsilon_1, \ldots, $ are i.i.d as $N(0,\sigma^2)$. Then for any pair $(\lambda, a)$ that satisfies $a>1/d_{\min}$, $\theta_{\star}^0>a\lambda$ and $NT\lambda^2 > \sigma^2$, we have
    $$
    P\left(\hat{\bftheta}(\lambda,a) =  \hat{\bftheta}^0\right) \geq 1- \psi_1(\lambda) - \psi_2(\lambda),
    $$
    where $\hat{\bftheta}^0=\underset{b}{\arg\min}\{\|\varY-\varX \mathbf{b}\|_2^2: \mathbf{b}_t=\mathbf{0}~\forall t\not\in S \}$ is the oracle least squares estimator.
\end{prop}
Proposition \ref{P5} guarantees an lower bound on the probability that the coefficient estimator with $l_2$ norm group MCP is identical to the oracle least squares estimator (though it is not a real estimator as the oracle set is unknown in practice). Further, define 
\begin{equation}\label{eq:lambdaN}
    \lambda_N=\sigma \left(2\log (\max\{T-|S|,1\}) / (NT g_{\min}(S^C))\right)^{1/2}
\end{equation}
\vspace{1ex} 
and
\vspace{1ex} 
\begin{equation}\label{eq:tauN}
    \tau_N=\sigma \sqrt{2\log\left( \max\{|S|,1\}\right) /(NT d_1 g_{\min}(S)) }.
\end{equation}
\vspace{1ex} 
Then the following corollary immediately follows.

\begin{corollary}
    Suppose that the conditions of Theorem 4 are satisfied. Assume that $\theta_{\star}^0 \geq a\lambda+a_N\tau_N$ for $a_N \rightarrow \infty$ as $N\rightarrow\infty$. If $\lambda \geq a_n\lambda_N$, then 
    $$
    P\left(\hat{\bftheta}(\lambda,a) =  \hat{\bftheta}^0\right) \rightarrow 1~\mbox{as}~N\rightarrow \infty.
    $$
\end{corollary}
The Corollary 1 shows that with probability tending to 1, the $l_2$ norm MCP estimator behaves like the oracle least squares estimator, and this implies the group selection consistent. 

\newpage
\section*{F. Additional numerical results}
\subsection*{F.1 Additional results for the scenario in the paper}

\begin{figure}[htp]
\centering
\includegraphics[scale=0.42]{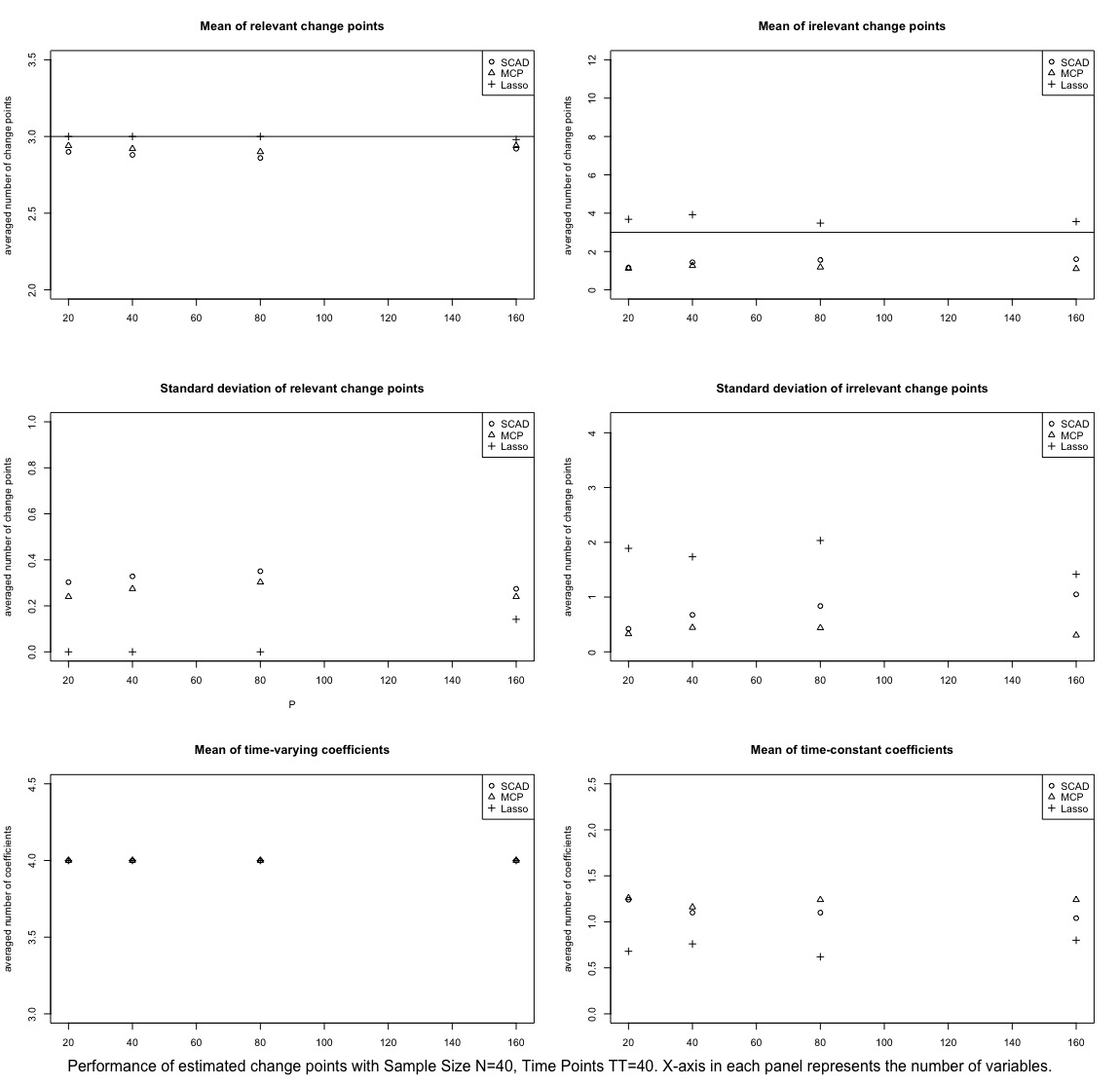}
\caption{Results of estimated change points with sample size N=40 and time points T=40.}
\label{N40TT40}
\end{figure}

\newpage
\begin{figure}[htp]
\center
\includegraphics[scale=0.42]{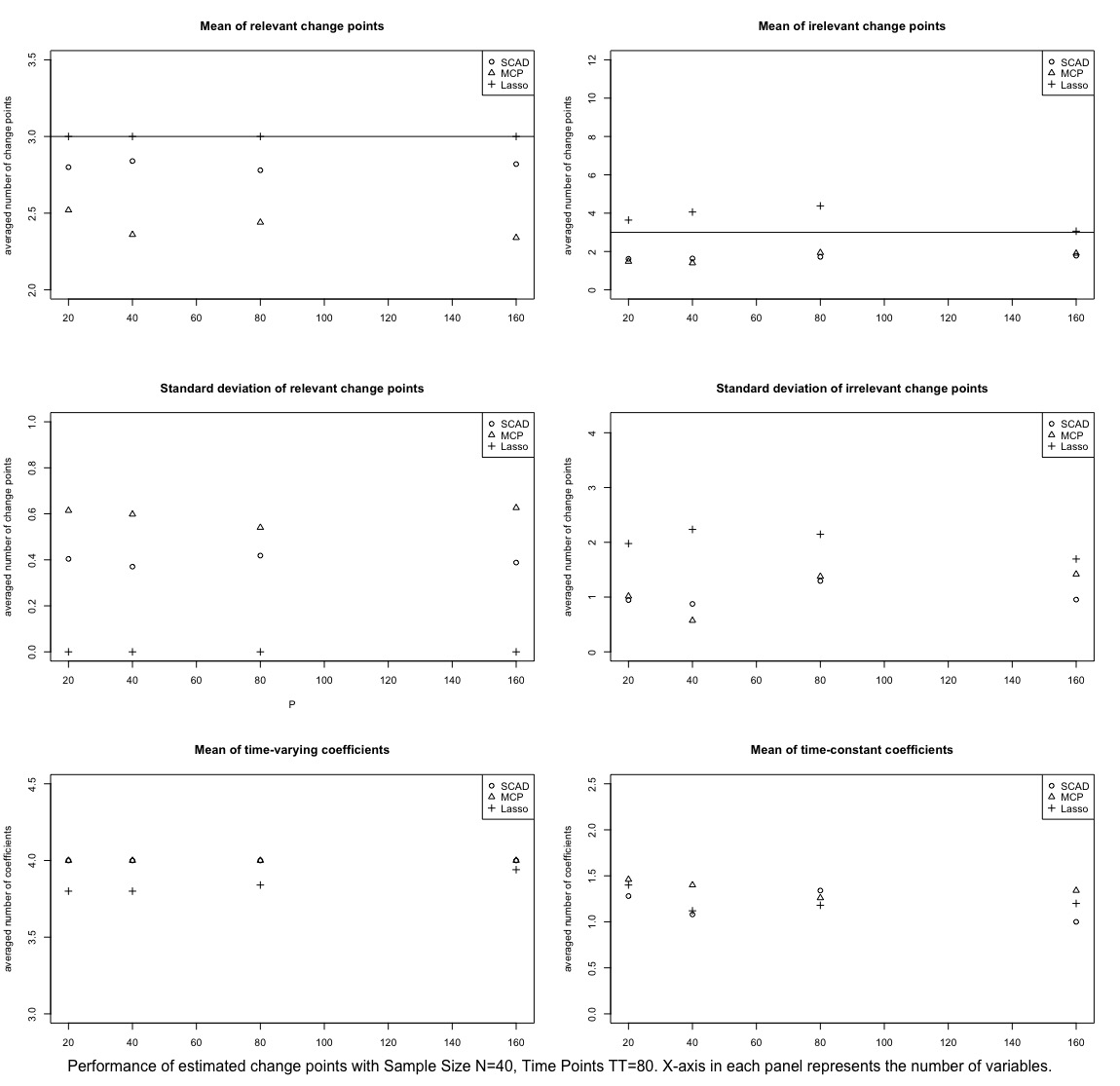}
\caption{Results of estimated change points with sample size N=40 and time points T=80.}
\label{N40TT80}
\end{figure}

\begin{figure}[htp]
\center
\includegraphics[scale=0.42]{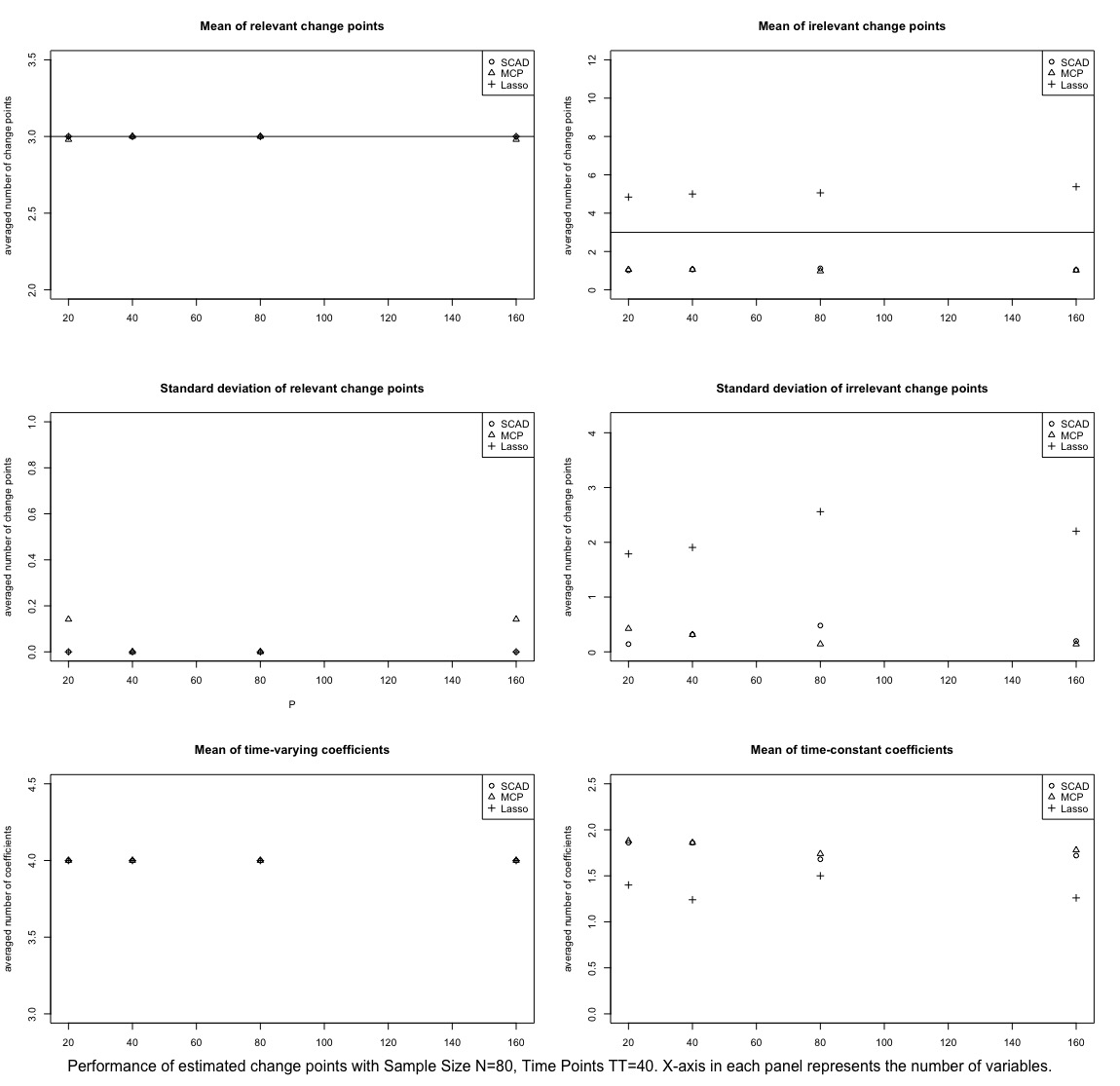}
\caption{Results of estimated change points with sample size N=80 and time points T=40.}
\label{N80TT40}
\end{figure}

\begin{figure}[htp]
\center
\includegraphics[scale=0.42]{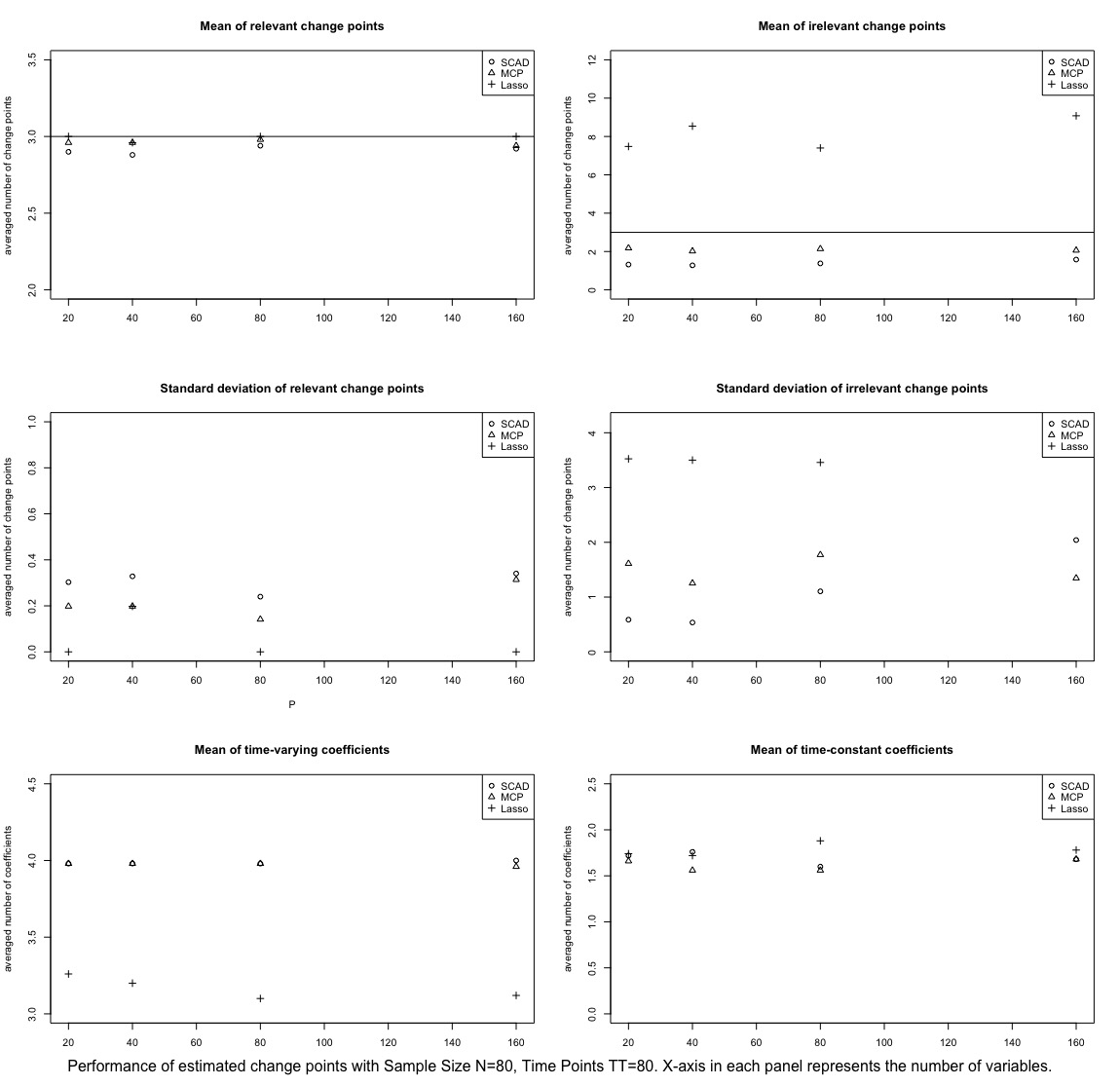}
\caption{Results of estimated change points with sample size N=80 and time points T=80.}
\label{N80TT80}
\end{figure}

\newpage
\begin{table}[htp]
\begin{center}
\caption{Results of the averaged of correctly selected change points and that of irrelevant change points from 100 repetitions, with sample size of 40, 80 and 200, and number of covariates of 20, 40, 80 and 160, respectively. The empirical standard deviation is in parentheses.}
\label{res_N}
\resizebox{\columnwidth}{!}{%
\begin{tabular}{lccccccccc}
\\
\toprule
& \multicolumn{3}{c}{\large{N=40}} & \multicolumn{3}{c}{\large{N=80}} & \multicolumn{3}{c}{\large{N=200}}\vspace{3pt}\\
\cmidrule(r){2-4}
\cmidrule(r){5-7}
\cmidrule(r){8-10}
 & SCAD & MCP & Lasso & SCAD & MCP & Lasso & SCAD & MCP & Lasso \vspace{3pt}\\ 
 \midrule
True Relevent (p=20) & 2.90 (0.30) & 2.94 (0.24) & 3.00 (0.00) & 3.00 (0.00) & 2.98 (0.14) & 3.00 (0.00) & 2.98 (0.14) & 2.98 (0.14) & 3.00 (0.00)\vspace{3pt}\\
Irrelevent (p=20) & 1.16 (0.42) & 1.12 (0.33) & 3.68 (1.89) & 1.02 (0.14) & 1.06 (0.42) & 4.84 (1.79) & 1.00 (0.00) & 1.00 (0.00) & 6.76 (2.58)\vspace{5pt}\\


True Relevent (p=40) & 2.88 (0.33) & 2.92 (0.27) & 3.00 (0.00) & 3.00 (0.00) & 3.00 (0.00) & 3.00 (0.00) & 2.98 (0.14) & 3.00 (0.00) & 3.00 (0.00)\vspace{3pt}\\
Irrelevent (p=40) & 1.44 (0.67) & 1.26 (0.44) & 3.92 (1.74) & 1.06 (0.31) & 1.06 (0.31) & 5.00 (1.91) & 1.00 (0.00) & 1.00 (0.00) & 6.66 (1.67)\vspace{5pt}\\


True Relevent (p=80) & 2.86 (0.35) & 2.90 (0.30) & 3.00 (0.00) & 3.00 (0.00) & 3.00 (0.00 )& 3.00 (0.00) & 3.00 (0.00) & 3.00 (0.00) & 3.00 (0.00)\vspace{3pt}\\
Irrelevent (p=80) & 1.56 (0.84) & 1.18 (0.44) & 3.48 (2.03) & 1.12 (0.48) & 0.98 (0.14) & 5.06 (2.56) & 1.00 (0.00) & 1.00 (0.00) & 7.38 (2.72)\vspace{5pt}\\


True Relevent (p=160) & 2.92 (0.27) & 2.94 (0.24) & 2.98 (0.14) & 3.00 (0.00) & 2.98 (0.14) & 3.00 (0.00) & 2.96 (0.20) & 2.98 (0.14) & 2.96 (0.20)\vspace{3pt}\\
Irrelevent (p=160) & 1.60 (1.05) & 1.10 (0.30) & 3.56 (1.42) & 1.04 (0.20) & 1.02 (0.14) & 5.38 (2.20) & 1.00 (0.00) & 1.00 (0.00) & 6.98 (2.71)\\
\bottomrule
\end{tabular}}
\end{center}
\end{table}


\begin{table}[htp]
\centering
\caption{Post selection for p=1600}
\resizebox{\columnwidth}{!}{
\begin{tabular}{lccccccccc}
 \\
\toprule
& \multicolumn{3}{c}{\large{N=40}} & \multicolumn{3}{c}{\large{N=80}} & \multicolumn{3}{c}{\large{N=160}}\vspace{3pt}\\
\cmidrule(r){2-4}
\cmidrule(r){5-7}
\cmidrule(r){8-10}
 & SCAD & MCP & Lasso & SCAD & MCP & Lasso & SCAD & MCP & Lasso \vspace{3pt} \\
 \midrule
True Relevent (TT=40) & 2.93(0.26) & 2.94(0.24) & 3.00(0.00) & 2.99(0.10) & 2.99(0.10) & 2.99(0.10) & 2.99(0.10) & 2.98(0.14) & 2.99(0.10) \\[3pt]
  Irrelevent (TT=40) & 1.18(0.41) & 1.20(0.45) & 3.99(2.08) & 1.00(0.14) & 1.00(0.00) & 4.91(1.89) & 1.00(0.00) & 1.00(0.00) & 6.74(2.33) \\[5pt]

  True Relevent (TT=80) & 2.80(0.47) & 2.44(0.67) & 2.98(0.14) & 2.89(0.31) & 2.98(0.14) & 3.00(0.00) & 2.98(0.14) & 2.99(0.10) & 3.00(0.00) \\[3pt]
  Irrelevent (TT=80) & 1.66(1.06) & 1.68(1.14) & 3.47(2.03) & 1.28(0.55) & 2.35(1.60) & 7.40(3.53) & 1.25(1.13) & 1.19(1.05) & 9.34(4.45) \\[5pt]

  True Relevent (TT=160) & 2.32(0.68) & 2.20(0.74) & 3.00(0.00) & 2.67(0.49) & 2.47(0.59) & 3.00(0.00) & 2.95(0.22) & 2.98(0.14) & 2.97(0.22) \\[3pt]
  Irrelevent (TT=160) & 1.54(0.83) & 2.10(1.40) & 4.56(2.02) & 3.16(3.62) & 1.93(1.11) & 8.96(2.77) & 2.95(2.10) & 6.87(5.94) & 13.98(3.41)\\
\bottomrule
\end{tabular}}
\end{table}

\begin{table}[htp]
\centering
\caption{Post selection outcomes for p=1600}
\resizebox{\columnwidth}{!}{
\begin{tabular}{lccccccccc}
 \\
\toprule
& \multicolumn{3}{c}{\large{N=40}} & \multicolumn{3}{c}{\large{N=80}} & \multicolumn{3}{c}{\large{N=160}}\vspace{3pt}\\
\cmidrule(r){2-4}
\cmidrule(r){5-7}
\cmidrule(r){8-10}
 & SCAD & MCP & Lasso & SCAD & MCP & Lasso & SCAD & MCP & Lasso \\[3pt]
 \midrule
  True Relevent (TT=40) & 2.53(0.52) & 2.54(0.52) & 2.17(0.77) & 2.29(0.46) & 2.29(0.46) & 1.46(0.63) & 2.15(0.36) & 2.15(0.36) & 1.10(0.30) \\[3pt]
  Irrelevent (TT=40) & 0.39(0.90) & 0.38(0.90) & 1.89(1.70) & -0.15(0.80) & -0.18(0.78) & 1.33(1.53) & -0.70(0.52) & -0.68(0.55) & 0.65(1.02) \\[5pt]
  
  True Relevent (TT=80) & 2.24(0.59) & 2.03(0.64) & 1.96(0.76) & 2.13(0.37) & 2.16(0.39) & 1.26(0.50) & 2.02(0.14) & 2.03(0.17) & 1.05(0.22) \\[3pt]
  Irrelevent (TT=80) & 0.15(1.10) & 0.22(1.35) & 0.82(1.41) & -0.63(0.63) & -0.10(1.24) & 0.67(1.15) & -0.87(0.37) & -0.84(0.44) & 0.07(0.41) \\[5pt]
  
  True Relevent (TT=160) & 1.75(0.50) & 1.73(0.51) & 1.62(0.65) & 1.83(0.45) & 1.78(0.48) & 1.13(0.39) & 1.98(0.14) & 2.01(0.17) & 1.00(0.00) \\[3pt]
  Irrelevent (TT=160) & -0.62(0.69) & -0.34(1.13) & 0.62(1.21) & -0.22(1.66) & -0.69(0.65) & 0.32(1.09) & -0.66(0.71) & -0.15(1.62) & 0.00(0.00)\\
   \bottomrule
\end{tabular}}
\end{table}

\begin{table}[htp]
\centering
\caption{p=1600,v2}
\resizebox{\columnwidth}{!}{
\begin{tabular}{lccccccccc}
    \\
    \toprule
    & \multicolumn{3}{c}{\large{N=40}} & \multicolumn{3}{c}{\large{N=80}} & \multicolumn{3}{c}{\large{N=160}}\vspace{3pt}\\
    \cmidrule(r){2-4}
    \cmidrule(r){5-7}
    \cmidrule(r){8-10}
 & SCAD & MCP & Lasso & SCAD & MCP & Lasso & SCAD & MCP & Lasso \\[3pt]
  \midrule
  True Relevent (TT=40) & 2.93(0.26) & 2.94(0.24) & 3.00(0.00) & 2.99(0.10) & 2.99(0.10) & 2.99(0.10) & 2.99(0.10) & 2.98(0.14) & 2.99(0.10) \\[3pt]
  Irrelevent (TT=40) & 1.15(0.39) & 1.16(0.42) & 3.60(1.93) & 1(0.14) & 1.00(0.00) & 4.45(1.79) & 1.00(0.00) & 1.00(0.00) & 6.28(2.21) \\[5pt]
  
  True Relevent (TT=80) & 2.80(0.47) & 2.44(0.67) & 2.98(0.14) & 2.89(0.31) & 2.98(0.14) & 3.00(0.00) & 2.98(0.14) & 2.99(0.10) & 3.00(0.00) \\[3pt]
  Irrelevent (TT=80) & 1.62(1.01) & 1.67(1.14) & 3.17(1.93) & 1.23(0.47) & 2.33(1.60) & 6.97(3.50) & 1.25(1.13) & 1.19(1.05) & 8.91(4.39) \\[5pt]
  
  True Relevent (TT=160) & 2.32(0.68) & 2.20(0.74) & 3.00(0.00) & 2.67(0.49) & 2.47(0.59) & 3.00(0.00) & 2.95(0.22) & 2.98(0.14) & 2.97(0.22) \\[3pt]
  Irrelevent (TT=160) & 1.54(0.83) & 2.09(1.39) & 4.19(1.99) & 3.15(3.62) & 1.92(1.12) & 8.49(2.75) & 2.93(2.08) & 6.87(5.94) & 13.40(3.45)\\
  \bottomrule
\end{tabular}}
\end{table}

\begin{table}[htp]
\centering
\caption{p=1600,v2,post}
\resizebox{\columnwidth}{!}{
\begin{tabular}{lccccccccc}
   \\
    \toprule
    & \multicolumn{3}{c}{\large{N=40}} & \multicolumn{3}{c}{\large{N=80}} & \multicolumn{3}{c}{\large{N=160}}\vspace{3pt}\\
    \cmidrule(r){2-4}
    \cmidrule(r){5-7}
    \cmidrule(r){8-10}
     & SCAD & MCP & Lasso & SCAD & MCP & Lasso & SCAD & MCP & Lasso \\[3pt]
     \midrule
     
  True Relevent (TT=40) & 2.53(0.52) & 2.54(0.52) & 2.17(0.77) & 2.29(0.46) & 2.29(0.46) & 1.46(0.63) & 2.15(0.36) & 2.15(0.36) & 1.10(0.30) \\[3pt]
  Irrelevent (TT=40) & 0.37(0.88) & 0.35(0.88) & 1.69(1.66) & -0.15(0.80) & -0.18(0.78) & 1.18(1.47) & -0.70(0.52) & -0.68(0.55) & 0.58(0.98) \\[5pt]
  
  True Relevent (TT=80) & 2.24(0.59) & 2.03(0.64) & 1.96(0.76) & 2.13(0.37) & 2.16(0.39) & 1.26(0.50) & 2.02(0.14) & 2.03(0.17) & 1.05(0.22) \\[3pt]
  Irrelevent (TT=80) & 0.12(1.06) & 0.21(1.34) & 0.68(1.39) & -0.63(0.63) & -0.11(1.24) & 0.61(1.10) & -0.87(0.37) & -0.84(0.44) & 0.06(0.34) \\[5pt]
  
  True Relevent (TT=160) & 1.75(0.50) & 1.73(0.51) & 1.62(0.65) & 1.83(0.45) & 1.78(0.48) & 1.13(0.39) & 1.98(0.14) & 2.01(0.17) & 1.00(0.00) \\[3pt]
  Irrelevent (TT=160) & -0.62(0.69) & -0.34(1.13) & 0.48(1.19) & -0.23(1.64) & -0.69(0.65) & 0.29(1.02) & -0.66(0.71) & -0.15(1.62) & 0.00(0.00)\\
   \bottomrule
\end{tabular}}
\end{table}


\section*{G. Additional simulation results for real data analysis}

\begin{figure}[htp]
    \centering
    \includegraphics[scale=0.7]{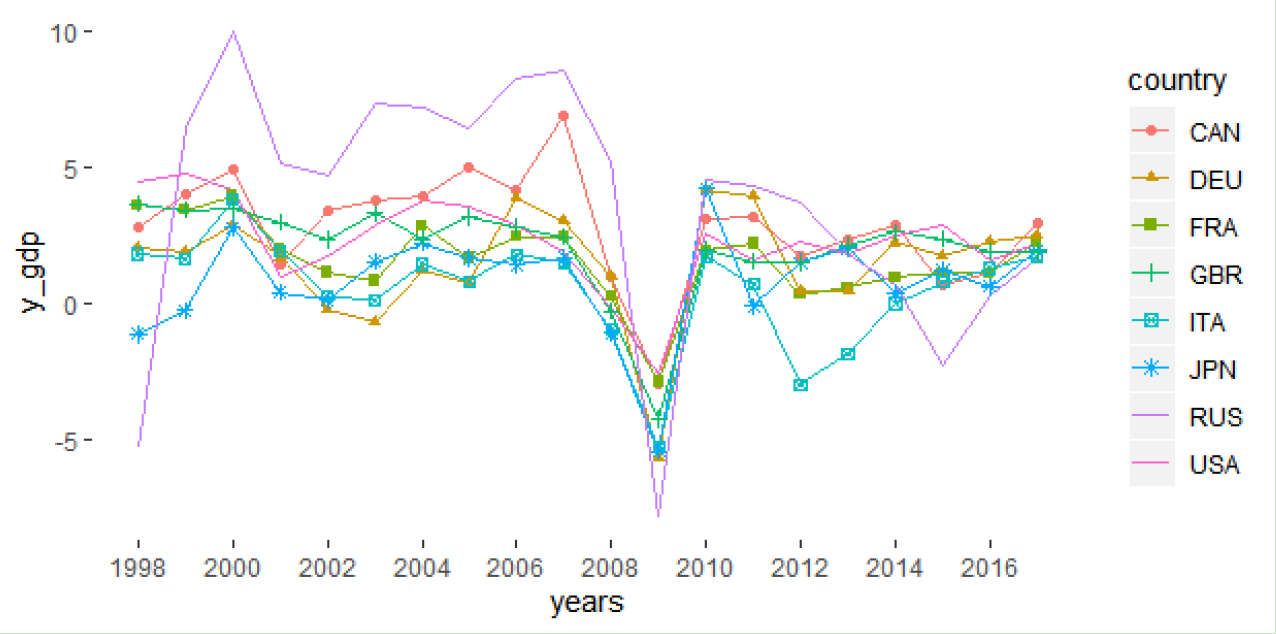}
    \caption{Scatter plot of G8 annual growth rates separately from 1998 to 2017.}
    \label{fig:gdp_allnation}
\end{figure}

\begin{table}[htp] 
    \centering 
	\caption{Descriptive statistics of G8 annual growth rates.} 
	\label{tab:des_stat}
	\renewcommand\tabcolsep{1.1pt} 
	\begin{tabular}{@{\extracolsep{5pt}} lllllllllllll} 
	\\
	\toprule
	group & n & mean & sd & median & trimmed & mad & min & max & range & skew & kurtosis & se \\ 
	\midrule
	1998 & 8 & 1.49 & 3.23 & 2.41 & 1.49 & 1.79 & -5.30 & 4.48 & 9.78 & -1.05 & -0.36 & 1.14 \\ 
	1999 & 8 & 3.15 & 2.05 & 3.42 & 3.15 & 2.12 & -0.25 & 6.40 & 6.65 & -0.09 & -1.16 & 0.73 \\ 
	2000 & 8 & 4.48 & 2.33 & 3.86 & 4.48 & 1.02 & 2.78 & 10.00 & 7.22 & 1.55 & 0.98 & 0.82 \\ 
	2001 & 8 & 2.06 & 1.44 & 1.82 & 2.06 & 0.92 & 0.41 & 5.10 & 4.69 & 0.95 & -0.25 & 0.51 \\ 
	2002 & 8 & 1.69 & 1.73 & 1.44 & 1.69 & 1.86 & -0.20 & 4.70 & 4.90 & 0.46 & -1.39 & 0.61 \\ 
	2003 & 8 & 2.38 & 2.54 & 2.19 & 2.38 & 2.21 & -0.71 & 7.30 & 8.01 & 0.59 & -0.84 & 0.90 \\ 
	2004 & 8 & 3.12 & 1.92 & 2.60 & 3.12 & 1.76 & 1.19 & 7.20 & 6.01 & 0.98 & -0.23 & 0.68 \\ 
	2005 & 8 & 2.87 & 2.04 & 2.42 & 2.87 & 2.00 & 0.72 & 6.40 & 5.68 & 0.47 & -1.43 & 0.72 \\ 
	2006 & 8 & 3.44 & 2.13 & 2.82 & 3.44 & 1.50 & 1.42 & 8.20 & 6.78 & 1.22 & 0.26 & 0.75 \\ 
	2007 & 8 & 3.53 & 2.65 & 2.43 & 3.53 & 0.99 & 1.49 & 8.50 & 7.01 & 0.93 & -1.02 & 0.94 \\ 
	2008 & 8 & 0.62 & 2.01 & 0.06 & 0.62 & 1.37 & -1.09 & 5.20 & 6.29 & 1.33 & 0.51 & 0.71 \\ 
	2009 & 8 & -4.60 & 1.80 & -4.76 & -4.60 & 2.05 & -7.80 & -2.54 & 5.26 & -0.35 & -1.32 & 0.64 \\ 
	2010 & 8 & 3.02 & 1.14 & 2.83 & 3.02 & 1.48 & 1.71 & 4.50 & 2.79 & 0.14 & -1.96 & 0.40 \\ 
	2011 & 8 & 2.16 & 1.55 & 1.87 & 2.16 & 1.81 & -0.12 & 4.30 & 4.42 & 0.04 & -1.61 & 0.55 \\ 
	2012 & 8 & 1.05 & 1.95 & 1.49 & 1.05 & 1.36 & -2.98 & 3.70 & 6.68 & -0.77 & -0.24 & 0.69 \\ 
	2013 & 8 & 1.16 & 1.40 & 1.82 & 1.16 & 0.61 & -1.84 & 2.33 & 4.17 & -1.08 & -0.26 & 0.50 \\ 
	2014 & 8 & 1.52 & 1.13 & 1.59 & 1.52 & 1.41 & 0.00 & 2.87 & 2.87 & -0.08 & -1.99 & 0.40 \\ 
	2015 & 8 & 1.06 & 1.56 & 1.17 & 1.06 & 0.78 & -2.31 & 2.88 & 5.19 & -0.98 & 0.01 & 0.55 \\ 
	2016 & 8 & 1.27 & 0.63 & 1.19 & 1.27 & 0.71 & 0.33 & 2.23 & 1.90 & 0.04 & -1.44 & 0.22 \\ 
	2017 & 8 & 2.14 & 0.44 & 2.07 & 2.14 & 0.40 & 1.63 & 2.98 & 1.35 & 0.59 & -1.00 & 0.16 \\ 
	\bottomrule
	\end{tabular} 
\end{table}

\begin{table} \centering
	\caption{Estimated useful variables from different screening methods.} 
	\label{tab:screen} 
	\begin{tabular}{@{\extracolsep{5pt}} c l} 
		\\
		\toprule
		Methods & Variables after being screened \\
	    \midrule
		\multirow{1}{*}{SIS} & Adjusted net total income per capita (annual \% growth) \\
		\midrule
		\multirow{3}{*}{HOLP} & Adjusted net total income (annual \% growth) \\ 
		&Adjusted net total income per capita (annual \% growth)\\ 
		&Gross fixed capital formation (annual \% growth) \\
		\midrule
		\multirow{10}{*}{DTCCS} & Adjusted net total income per capita (annual \% growth)  \\ 
		&Employment in agriculture (\% of total employment) (modeled ILO estimate) \\ 
		&Employment in industry (\% of total employment) (modeled ILO estimate)\\ 
		&Employment in services (\% of total employment) (modeled ILO estimate) \\ 
		&Employment to population ratio, ages 15-24, total (\%) (modeled ILO estimate) \\ 
		&Households and NPIs Non-Financial consumption expenditure (annual \% growth)\\ 
		&Labor force participation rate for ages 15-24, total (\%) (total estimate)\\ 
		&Population ages 15-64, total \\ 
		&Population in largest city\\ 
		&Population in urban agglomerations of more than 1 million (\% of total population)  \\
		\bottomrule
	\end{tabular} 
\end{table}

\newpage
\bibliographystyle{chicago}
\bibliography{dtccs-cpt.bib}